\documentclass[iop]{emulateapj}
\usepackage{epsfig}
\usepackage{amsmath}
\usepackage{natbib}
\usepackage{amsmath}
\usepackage{graphicx}
\usepackage{lineno}
\usepackage{color}
\usepackage[colorlinks=true,
            linkcolor=blue,
            urlcolor=black,
            citecolor=blue]{hyperref}

\shorttitle{{\it Fermi}-LAT GRBs as cosmological standard candles}
\shortauthors{Dirirsa, Razzaque \& Piron}                        

\begin{document}
\title{Spectral analysis of \textit{Fermi}-LAT gamma-ray bursts with known redshift and their potential use as cosmological standard candles}


\author{
F.~Fana~Dirirsa\altaffilmark{1,2},
S.~Razzaque\altaffilmark{1,3}, 
F.~Piron\altaffilmark{4,5}, 
M.~Arimoto\altaffilmark{6}, 
M.~Axelsson\altaffilmark{7,8}, 
D.~Kocevski\altaffilmark{9}, 
F.~Longo\altaffilmark{10,11}, 
M.~Ohno\altaffilmark{12}, 
S.~Zhu\altaffilmark{13}
}
\altaffiltext{1}{Department of Physics, University of Johannesburg, PO Box 524, Auckland Park 2006, South Africa}
\altaffiltext{2}{email: fdirirsa@uj.ac.za}
\altaffiltext{3}{email: srazzaque@uj.ac.za}
\altaffiltext{4}{Laboratoire Univers et Particules de Montpellier, Universit\'e Montpellier, CNRS/IN2P3, F-34095 Montpellier, France}
\altaffiltext{5}{email: piron@in2p3.fr}
\altaffiltext{6}{Faculty of Mathematics and Physics, Institute of Science and Engineering, Kanazawa University, Kakuma, Kanazawa, Ishikawa 920-1192}
\altaffiltext{7}{Department of Physics, KTH Royal Institute of Technology, AlbaNova, SE-106 91 Stockholm, Sweden}
\altaffiltext{8}{Tokyo Metropolitan University, Department of Physics, Minami-osawa 1-1, Hachioji, Tokyo 192-0397, Japan}
\altaffiltext{9}{NASA Goddard Space Flight Center, Greenbelt, MD 20771, USA}
\altaffiltext{10}{Istituto Nazionale di Fisica Nucleare, Sezione di Trieste, I-34127 Trieste, Italy}
\altaffiltext{11}{Dipartimento di Fisica, Universit\`a di Trieste, I-34127 Trieste, Italy}
\altaffiltext{12}{Department of Physical Sciences, Hiroshima University, Higashi-Hiroshima, Hiroshima 739-8526, Japan}
\altaffiltext{13}{Albert-Einstein-Institut, Max-Planck-Institut f\"ur Gravitationsphysik, D-30167 Hannover, Germany}

\begin{abstract}
Long duration Gamma-Ray Bursts (LGRBs) may serve as standard candles to constrain cosmological parameters by probing the Hubble diagram well beyond the range of redshift currently accessible using type-Ia supernovae. The standardization of GRBs is based on phenomenological relations between two or more parameters found from spectral modeling, of which one is strongly dependent on the cosmological model. The Amati relation links the source-frame energy  ${E_{\mathrm{i,p}}}$ at which the prompt gamma-ray spectral energy distribution $\nu F_\nu$ peaks, and the isotropic-equivalent bolometric energy ${E_{\mathrm{iso}}}$ emitted during the prompt phase. We performed spectral analysis of 26 GRBs with known redshift that have been detected by the {\it Fermi}-Large Area Telescope (LAT) during its nine years of operations from July 2008 to September 2017, thus extending the computation of ${E_{\mathrm{iso}}}$ to the 100 MeV range. Multiple components are required to fit the spectra of a number of GRBs. We found that the Amati relation is satisfied by the 25 LGRBs, with best fit parameters similar to previous studies that used data from different satellite experiments, while the only short GRB with known redshift is an outlier. Using the Amati relation we extend the Hubble diagram to redshift 4.35 and constrain the Hubble constant and dark-energy density in the $\Lambda$CDM model, with {\it Fermi}-LAT GRBs alone and together with another sample of 94 GRBs and with the latest Supernovae type-Ia data. Our results are consistent with the currently acceptable ranges of those cosmological parameters within errors.
\end{abstract}


\section{Introduction}
Gamma-Ray Bursts (GRBs) are among the most energetic astronomical events whose power is emitted within a short period of time and is dominant in the (sub-)MeV gamma-ray range \citep{Klebesadel1973_182, Fishman1995}. The duration of a GRB is determined from the photon flux accumulation over time, typically between 5$\%$ and 95$\%$ of the fluence, and is called ${T_{90}}$ \citep{2012ApJS..199...18P}.  Based on their duration, GRBs are categorized between long ($T_{90} \gtrsim 2$~s) and short ($T_{90} \lesssim 2$~s) classes \citep{Kouveliotou1993_413}.  The progenitors of these two classes are thought to be different; the long GRBs (LGRBs) are results of core collapse of massive stars \citep{MacFadyen1999} while  short GRBs (SGRBs) are results of binary mergers of compact objects \citep{Eichler1989}.  Observational evidence, namely association of supernovae (SNe) with LGRBs \citep{Kulkarni1998, Stanek2003, Soderberg2006} and association of gravitational waves with an SGRB \citep{GW1708172017, GRB170817A} support these progenitor theories.      

GRBs are cosmological events and have been detected up to very high redshift of $z\sim 9.4$ \citep{Cucchiara2011_736}. If they can be standardized, similarly to type Ia SNe \citep{Riess1998_116, Perlmutter1999_517, Perlmutter_195}, GRBs could potentially be used as cosmological probes of the distant Universe. 

GRBs can be standardized based on phenomenological correlations between the observed spectral parameters and energetics. The Amati relation \citep{Amati2002_81A, Amati2006_121, Amati2006_372, Amati2008_391, Amati2009_508} is between the isotropic-equivalent radiated energy $E_{\rm iso}$ and the redshift-corrected energy $E_{\rm i,p}$ at which the time-averaged $\nu F_{\nu}$ spectrum peaks. It has been the most studied relation for LGRBs so far. Another important empirical relation between the $E_{\rm i,p}$ and the intrinsic peak luminosity ($L_{\rm iso}$) was discovered by \cite{Yonetoku2004_609}, which is followed by both the LGRBs and SGRBs \citep{Ghirlanda2009_496}. Yet another phenomenological correlation has been studied \citep{Ghirlanda2004_616} between $E_{\rm i,p}$ and the collimation corrected true emitted energy ($E_{\rm \gamma}$). More recently, \cite{Guiriec2013_770} also proposed a relation between the time-resolved luminosity $L_\mathrm{i}^{\rm Band}$ and $E_{\rm i,p}^{\rm rest}$. This relation holds the potential to determine redshifts using only  the study of the spectral evolution in the gamma-ray emission of GRBs.  Only the LGRBs follow the Amati relation and no convincing physical explanation is known.  The SGRBs are known to be inconsistent with the $E_\mathrm{i,p}$--$E_\mathrm{iso}$ correlation for LGRBs as explored by \citet{Ghirlanda2009_496}.  
Selection effects including detector artifacts may also play a significant role in the Amati relation \citep{Butler2007, Li2007, Ghirlanda2008_387, Butler2009, Butler2010, Collazzi2012, Kocevski2012, 2015ApJ...806...44P}. \cite{Heussaff2013_557A} found that this relation was partially due to a true lack of luminous GRBs with low $E_{\rm i,p}$ at the upper left boundary of the relation (see Fig.\,1 therein), and partially shaped by selection effects in the lower right part of the ($E_{\rm i,p}$, $E_{\rm iso}$) plane, namely by the limited efficiency of gamma-ray instruments for detecting GRBs with large $E_{\rm i,p}$ and low $E_{\rm iso}$. The amplitude of these instrumental effects is difficult to quantify owing to the incomplete knowledge of the underlying GRB population. However, they may not prevail if one instead considers the time-resolved luminosity-hardness relation ($E_{\rm i,p}^{\rm rest}$, $L_{\rm i}^{\rm Band}$) that has been proposed by \cite{Guiriec2013_770}. 
Nevertheless, the  strong ${E_{\rm i,p}}$--${E_{\rm iso}}$ correlation found in several studies has not precluded their use as cosmological standard candles.  
 
In this paper we analyze GRB data obtained with the \textit{Fermi} Gamma-ray Space Telescope during the 2008--2017 period and to test the Amati relation.  Subsequently we apply the Amati relation to constrain the Hubble constant $H_0$ and dark energy parameter $\Omega_\Lambda$ in a flat $\Lambda$CDM cosmological model.  The GRB Monitor \citep[GBM,][]{Meegan2009_702} and Large Area Telescope \citep[LAT,][]{Atwood2009_697} onboard \textit{Fermi} cover an energy range from 8 keV to above 300 GeV.  The 12 NaI detectors of GBM are sensitive in the 8 keV--1 MeV range, while the 2 BGO detectors cover 200 keV--40 MeV range.  The LAT detects photons with energy from 20 MeV to over 300 GeV.  Observation of GRBs with \textit{Fermi} has enabled us to characterize the broad-band prompt emission, revealing a spectral diversity and the presence of multiple spectral components. The empirical Band function \citep{Band1993_413}, which consists of two smoothly joined power laws, has been widely used in the past to represent the prompt emission spectrum of most GRBs in the keV-MeV energy range. An additional power law (PL) function was introduced to fit the spectrum of a long GRB \citep{Gonzalex2003_424} detected by the \textit{Compton Gamma-Ray Observatory} (CGRO) and to fit the spectrum of a short GRB \citep{Ackermann2010_716} detected by \textit{Fermi}. The presence of a thermal black body (BB) component in addition to a smoothly broken power law (SBPL) function was investigated in the CGRO data \citep{Ryde1999_39}, and in addition to the Band function in the \textit{Fermi} data  \citep{Guiriec2011_727}. More complex scenarios involving three models to fit spectra have also been investigated \citep{Guiriec2015_807, Guiriec2016_819}.
An overview of the possible physical interpretations of these features can be found, \citep[e.g., in][for an overview]{Gehrels2013}.    

We selected the GRBs simultaneously detected by \textit{Fermi} GBM and LAT with known redshift for which the spectral fits are well-constrained within the $T_{\rm 90}$ duration of the GBM. This forms a sample of 25 LGRBs and 1 SGRB with an exceptional spectral coverage. We combined the GBM data with the LAT Low Energy data ($\sim$ 30 - 100 MeV, see \cite{Pelassa_2010}) and LAT Pass 8 data (above 100 MeV) in a joint spectral analysis considering various combinations of spectral components. This allowed to constrain the intrinsic peak energy $E_{\rm i,p}$ = $E_{\rm p} (1 + z)$ of the GRB spectral energy distribution (SED) and to derive their isotropic equivalent energy $E_{\rm iso}$ from keV energies up to 100 MeV.

In addition to the \textit{Fermi} GRB sample, we reanalyze the GRB sample that \cite{Wang2016_585} used to fit the Amati relation. We compare and contrast our fits to the Amati relation with fits for this sample as well as from other work \citep{Heussaff2013_557A,  Demianski2017_693}. We also analyze the Amati relation for the joint samples of GRBs and for GRBs in two different redshift bins below and above $z$ = 1.414 (i.e., the maximum redshift value of the SNe U2.1 sample \citep{Suzuki2012_746}). Finally we use a simple analysis to explore possibilities of constraining the cosmological parameters $H_0$ and $\Omega_\Lambda$ with our \textit{Fermi} GRB sample and with various combinations of data, including the recent type Ia SNe sample \citep{Suzuki2012_746}.  

This paper is organized as follows: in Section \S\ref{sec_2} we discuss the sample selection and data analysis criteria; in Section \S\ref{sec_3} we perform time-integrated spectral analysis over $T_{90}$ duration, calculate ${E_{\rm iso}}$ and ${E_{\rm i,p}}$ for the selected GRBs; in Section \S\ref{sec_4} we perform Amati relation fits to our data and to other joint GRB samples then we carry out a cosmological analysis using our parametrized Amati relation in Section \S\ref{sec_5}; and finally we discuss our results in Section \S\ref{sec_6}.

\section{Data sets}\label{sec_2} 
\subsection{GRB samples}
From the launch of \textit{Fermi} on 11 June 2008 until September 2017, about 32 GRBs were detected with identified redshift $z$, including the marginally detected LAT GRB\,091208B \citep{2013LATGRBCAT}. Of these, GBM did not trigger on GRB\,081203A and GRB\,130907A. Also, we did not find sufficient LAT photons within GBM $T_{\rm 90}$ for GRB\,160623A, GRB\,130702A and GRB\,120711A to perform a time-integrated joint spectral analysis of the LAT and GBM data. By excluding these six GRBs, we conduct our analysis only for the \textit{Fermi} GRB sample of 25 LGRBs and 1 SGRB with well-constrained spectral properties. This covers from GRB 080916C to GRB 170405A, as listed in Table~\ref{tab:modelfit}. The spectroscopic or photometric redshifts of these GRBs have been obtained from various GCN notices and published papers, as noted in Table~\ref{tab:Sbolo}. In addition to the \textit{Fermi} GRB sample, we also use a sample of 94 GRBs from \citet{Wang2016_585}, hereafter W2016, for comparisons.  These 94 GRBs have been selected from 151 GRBs analyzed by \citet{Wang2016_585}, and which are not in our \textit{Fermi} GRB sample or counted twice. We update the relevant GRB parameters of the W2016 sample using the latest cosmological model, see Appendix~\ref{W26_updated}. In particular, unless otherwise stated, we have adopted $\Lambda$CDM cosmology with standard parameters \citep{Bennett2014_794, Planck2018}, the dark energy density, $\Omega_{\mathrm{\Lambda}} = 0.714$, total density of baryonic and dark matter, $\Omega_{\mathrm{m}} = 0.286$,  the Hubble parameter $H_{\mathrm{0}} = 69.6$ km s$^{-1}$ Mpc$^{-1}$  and a spatially flat Universe.    

\begin{figure}[htp]
\begin{center}
\includegraphics[width=7.9 cm]{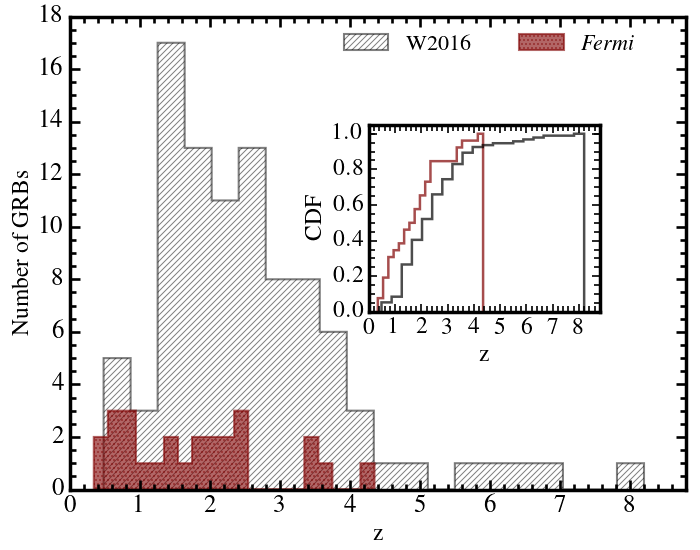}
\includegraphics[width=7.9 cm]{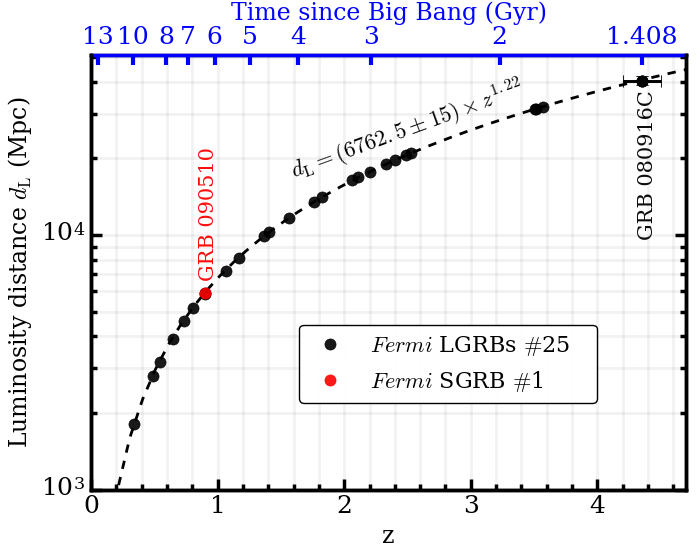}
\caption{\textit{Top panel --} Redshift distribution for our sample of 26 (25 long and 1 short) {\it Fermi} GRBs (maroon histogram) and the W2016 sample of 94 GRBs (grey histogram).  The cumulative distribution function (CDF) of the {\it Fermi} (median: $z$ = 1.66, mean: $z$ = 1.79) and  W2016 (median: $z$ = 2.30, mean: $z$ = 2.57) samples are indicated in the inset with maroon and grey lines, respectively.  \textit{Bottom panel --} The luminosity distance $d_L$ of 26 {\it Fermi} GRBs in $\Lambda$CDM cosmology with standard parameters vs. redshift.  The upper $x$-axis shows the time (Gyr) since Big Bang.}
\label{tab:histogram}
\end{center}
\end{figure}

In Fig.~\ref{tab:histogram} we show the redshift distribution of our \textit{Fermi} GRB sample and that of W2016. Unlike the redshift of GRBs peaking at $\approx 1.5$ for the W2016 sample, \textit{Fermi} GRBs are rather uniformly distributed in redshift and are comparatively closer. This could be due to a selection effect and to the smaller size of the \textit{Fermi} sample. The bottom panel of Fig.~\ref{tab:histogram} shows the luminosity distances of \textit{Fermi} GRBs with standard parameters of the $\Lambda$CDM cosmology.  The closest one is GRB\,130427A at $z=0.3399$. 

\subsection{{\it Fermi} data preparation}
We have implemented the criteria in \citet{Guiriec2011_727} for GBM detector selection from \textit{Fermi} data, as discussed below. We used the \texttt{rmfit} package\footnote{https://fermi.gsfc.nasa.gov/ssc/data/analysis/rmfit/} to simultaneously fit the spectral data of GBM NaI detectors that have source incidence angles smaller than $50^\circ$. In order to subtract background in GBM data, we fitted a second-order polynomial to data collected from two-time intervals selected before and after the prompt emission. Then this background model has been interpolated across the source selection time interval. We have also used the standard 128 energy bins of the \texttt{CSPEC} data-type, for NaI using the channels from $\sim$ 8 keV to $\sim$ 900 keV by cutting out the overflow high energy channels as well as the Iodine K-edge from $\sim$ 30 to $\sim$ 40 keV \citep{Meegan2009_702}. For the GBM BGO detectors, we have used data from $\sim$ 220 keV to $\sim$ 40 MeV and from $\sim$ 210 keV to $\sim$ 40 MeV, respectively for detectors \texttt{b0} and \texttt{b1} \citep{2016A&A...588A.135Y}.  We use 2 BGO detectors when the source angle is less than 100 degrees for both and in addition, NaI detectors from both the 0--5 and 6--11 groups are triggered. The detectors selected for each GRB are listed in Table~\ref{tab:modelfit}.

For the analysis of LAT data, we have selected the Pass 8 Transient class events (\texttt{Transient20E})\footnote{https://fermi.gsfc.nasa.gov/ssc/data/analysis/documentation/\\Cicerone/Cicerone$\_$Data/LAT$\_$DP.html} within a 10$^\circ$ radius of interest. The data is binned in 30 logarithmic energy steps between 30 MeV and 300 GeV.  Since we have considered energies below 100 MeV, the \texttt{gtlike}\footnote{https://fermi.gsfc.nasa.gov/ssc/data/analysis/scitools/help/\\gtlike.txt} tool was used to perform a binned maximum-likelihoood analysis that includes a correction for the energy dispersion effect. We produced the observed spectrum and the detector response matrix using the \textit{Fermi} Science Tools \texttt{gtbin}\footnote{https://fermi.gsfc.nasa.gov/ssc/data/analysis/scitools/help/\\gtbin.txt} and \texttt{gtrspgen}\footnote{https://fermi.gsfc.nasa.gov/ssc/data/analysis/scitools/help/\\gtrspgen.txt}, respectively. To produce a background spectrum file, the background estimation tool \texttt{gtbkg}\footnote{https://fermi.gsfc.nasa.gov/ssc/data/analysis/scitools/help/\\gtbkg.txt} was used.

\begin{table*}[htb!]
\begin{center}
\caption{\textit{Fermi} sample of 25 GRBs with selected detectors and results from the spectral model fits of time-integrated flux within $T_{90}$.}
\scalebox{0.70}{
\begin{tabular}{@{}l*{45}{l}}  
\hline\hline 
GRB name    & detectors                  & model      & $T_{05}-T_{95}$ (s)        & $\alpha$, $\gamma$         & $\beta$          & ${E_\mathrm{p}}$  (keV) & $kT$ (keV)        & $\alpha_1$       & C-Stat/dof$^{(*)}$\\
\hline
GRB\,170405A & n6+n7+n9+nb+b1+LAT    & Band         & 7.36-86.08      & -0.84 $\pm$ 0.01   & -2.44 $\pm$ 0.02 & 315.8 $\pm$ 7.78     &                  &                  & 1544.1/588         \\
GRB\,170214A                                           & n0+n1+n3+b0+LAT       & SBPL+BB      & 12.54-135.49    & -1.17 $\pm$ 0.02   & -2.51 $\pm$ 0.01 & 507.7 $\pm$ 34.9     & 41.99 $\pm$ 1.28 &                  & 1253.7/368         \\
GRB\,160625B                                            & n7+n9+b1+LAT          & Band+BB+PL   & 188.45-650.54   & -0.40 $\pm$ 0.06   & -2.70 $\pm$ 0.02 & 642.92 $\pm$ 15.48   & 27.94 $\pm$ 1.09 & -2.16 $\pm$ 0.04 & 1462.9/354         \\
GRB\,160509A                                            & n0+n1+n3+b0+LAT       & Band+CPL+PL & 7.68-379.4      & -0.87 $\pm$ 0.08   & -5.16 $\pm$ 0.49 & 8591.48 $\pm$ 68.27  &                  &                  &                    \\
                                                       &                       &              &                 & -0.79 $\pm$ 0.04   &                  & 317.14 $\pm$ 16.60   &                  & -1.76 $\pm$ 0.10 & 1741.9/474         \\
GRB\,150514A                                            & n3+n6+n7+b0+LAT       & Band         & 0.00-10.8       & -1.45 $\pm$ 0.08   & -2.33 $\pm$ 0.05 & 76.28 $\pm$ 8.26     &                  &                  & 590.57/472         \\
GRB\,150403A                                            & n3+n4+b0+LLE          & Band+BB      & 3.33-25.60      & -1.02 $\pm$ 0.02   & -2.95 $\pm$ 0.10 & 793.63 $\pm$ 52.55   & 33.30 $\pm$ 1.58 &                  & 524.75/358         \\
GRB\,150314A                                            & n0+na+n1+n9+b1+LAT    & Band         & 0.6-11.29       & -0.63 $\pm$ 0.01   & -3.02 $\pm$ 0.10 & 357.38 $\pm$ 4.78    &                  &                  & 1333.0/588         \\
GRB\,141028A                                            & n6+n7+n9+b1+LAT       & Band         & 6.66-38.16      & -0.91 $\pm$ 0.02   & -2.37 $\pm$ 0.02 & 396.45 $\pm$ 15.29   &                  &                  & 691.79/473         \\
GRB\,131231A                                            & n0+n3+n7+b0+LAT       & Band         & 13.31-44.31     & -1.23 $\pm$ 0.01   & -2.65 $\pm$ 0.03 & 225.17 $\pm$ 3.02    &                  &                  & 1665.0/476         \\
GRB\,131108A                                            & n0+n3+n6+n7+b0+b1+LAT & SBPL         & 0.32 - 19.32    & -0.99 $\pm$ 0.02   & -2.23 $\pm$ 0.01 & 205.32 $\pm$ 6.91    &                  &                  & 950.58/716         \\
GRB\,130518A                                            & n3+n6+n7+b0+b1+LAT    & Band         & 9.9-57.9        & -0.89 $\pm$ 0.01   & -2.71 $\pm$ 0.03 & 458.85 $\pm$ 9.22    &                  &                  & 1357.1/592         \\
GRB\,130427A                                            & n6+n9+na+b1+LAT       & Band+PL      & 11.23-142.34    & -1.41 $\pm$ 0.01   & -2.27 $\pm$ 0.01 & 219.61 $\pm$ 4.38    &                  & -1.22 $\pm$0.21  & 2105.1/488         \\
GRB\,120624B                                            & n1+n2+na+b0+b1+LAT    & SBPL         & -258.05-13.31   & -1.04 $\pm$ 0.01   & -2.78 $\pm$ 0.04 & 352.9 $\pm$ 11.4     &                  &                  & 2015.7/588         \\
GRB\,110721A               & n6+n7+n9+b1+n11+LAT       & Band+BB         & 0.45 - 24.9     & -1.24 $\pm$ 0.01   & -2.89 $\pm$ 0.06 & 1923.0 $\pm$ 189.0   &       34.05 $\pm$ 1.58           &                  & 770.32/586         \\
GRB\,100728A                                           & n0+n1+n2+n5+b0+LAT    & Band         & 13.25-178.75    & -0.52 $\pm$ 0.02   & -2.63 $\pm$ 0.04 & 310.7 $\pm$ 7.06     &                  &                  & 3075.3/595         \\
GRB\,100414A                                            & n7+n9+n11+b1+LAT      & Band         & 2.0 - 28.4      & -0.50 $\pm$ 0.02   & -2.91 $\pm$ 0.06 & 578.89 $\pm$ 11.69   &                  &                  & 750.82/469         \\
GRB\,091208B                                            & n10+n9+b1+LAT         & Band         & 0.26 - 15.26    & -1.29 $\pm$ 0.07   & -2.53 $\pm$ 0.12 & 98.22 $\pm$ 9.74     &                  &                  & 422.09/351         \\
GRB\,091127                  & n6+n7+n9+b1+LAT       & SBPL      & 0.00-7.80       & -1.42 $\pm$ 0.05   & -2.33 $\pm$ 0.02 & 33.10  $\pm$ 2.38    &  &                  & 731.85/479         \\
GRB\,091003A                                            & n0+n3+n6+b0+b1+LAT    & Band         & 1.09 - 22.19    & -1.08 $\pm$ 0.01   & -2.79 $\pm$ 0.05 & 452.21 $\pm$ 17.44   &                  &                  & 674.54/600         \\
GRB\,090926A   & n6+n7+n8+b1+LAT       & Band+PL         & 2.05-22.05      & -0.66 $\pm$ 0.03   & -2.34 $\pm$ 0.02 & 279.60 $\pm$ 4.51    &                  &      -1.82 $\pm$ 0.03            & 918.19/476         \\
GRB\,090902B                                            & n0+n2+n9+b0+b1+LAT    & Band+PL      & 0-22            & -0.53 $\pm$ 0.01   & -4.14 $\pm$ 0.28 & 760.66 $\pm$ 7.69    &                  & -1.92 $\pm$ 0.01 & 1320.6/601         \\
\textbf{GRB\,090510}        & \textbf{n3+n6+n7+n9+b0+b1+LAT} & \textbf{Band+PL}         & \textbf{0.002-1.744}     & \textbf{-0.63 $\pm$ 0.08}   & \textbf{-2.57 $\pm$ 0.08} & \textbf{3805.0 $\pm$ 385.0} &                  &         \textbf{-1.60 $\pm$ 0.03}         & \textbf{756.57/717}         \\
GRB\,090424                 & n7+n8+nb+b1+LAT       & Band      & 0.448-14.720    & -0.83 $\pm$ 0.02   & -2.49 $\pm$ 0.04 & 153.4 $\pm$ 2.91     &  &                  & 857.75/474         \\
GRB\,090328                                             & n7+n8+b1+LAT          & Band         & 4.67-61.67      & -1.04 $\pm$ 0.02   & -2.37 $\pm$ 0.04 & 703.75 $\pm$ 47.16   &                  &                  & 769.08/360         \\
GRB\,090323                                             & n6+n7+n9+n11+b1+LAT   & SBPL         & -1.0 - 173      & -1.29 $\pm$ 0.01   & -2.50 $\pm$ 0.02 & 399.44 $\pm$ 17.17   &                  &                  & 1558.6/597         \\
GRB\,080916C   & n3+n4+b0+LAT          & Band+BB         & 0 - 66          & -1.27 $\pm$ 0.03   & -2.28 $\pm$ 0.03 & 1297.0 $\pm$ 222.0   &   46.78  $\pm$ 1.93               &                  & 536.28/362 
\\  
\hline
\label{tab:modelfit}
\end{tabular}
}
\end{center}
\scriptsize{\textbf{Notes:} ${\alpha}$ and ${\beta}$ are the lower and higher photon indices for the Band and SBPL functions, respectively.   ${\gamma}$ is the photon index of CPL model while $\alpha_{\rm 1}$ is that of the PL.  $E_{\rm 0}$ is the SBPL e-folding energy and $E_{\rm p}$ is the Band or CPL peak energy. $kT$ is the BB temperature. The C-Stat/dof$^{(*)}$ is the ratio of the C-stat resulting from the fit and the associated degrees of freedom (dof).  GRB\,090510 is the only short GRB in the sample.}
\end{table*}

\section{Spectral analysis}\label{sec_3}
\subsection{Spectral fitting}
\begin{figure*}[htb!]
\begin{center}
\includegraphics[width=14 cm]{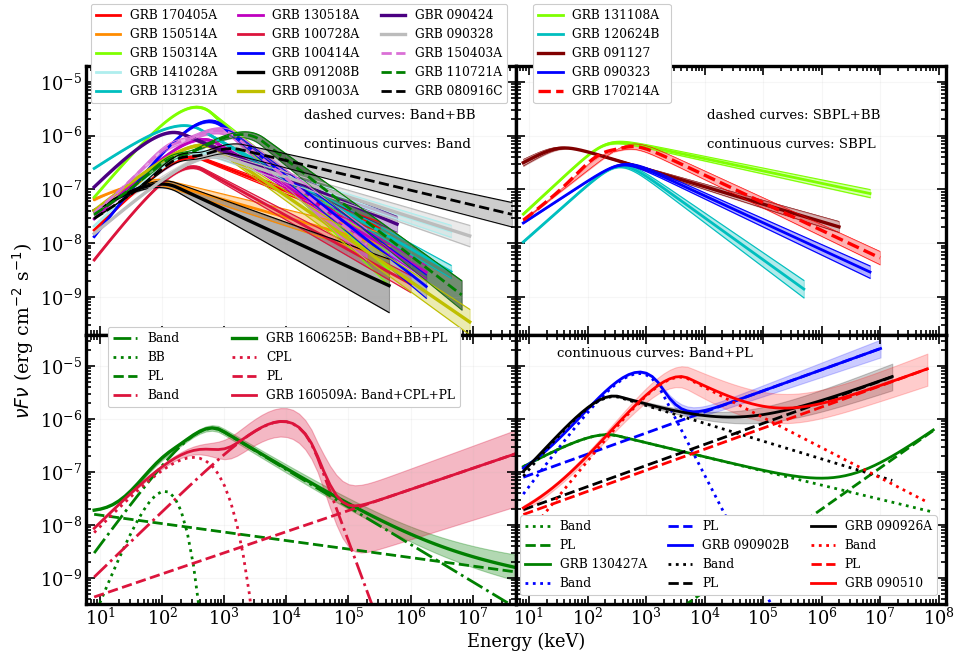}
\caption{$\nu F_{\nu}$ spectra of {\it Fermi}-LAT/GBM GRBs with known redshift resulting from the time integrated spectral analysis over the $T_{90}$ duration. The top left panel shows the Band or Band+BB model fits, while the top right panel shows SBPL or SBPL+BB model fits.  The bottom panels show more complex models along with their components.  The $1\sigma$ confidence regions of the models are shown with shades of the same color as the model lines.}
\label{fig:models}
\end{center}
\end{figure*}

In order to model the spectra of GRB prompt emission, we have performed a spectral analysis over the $T_{\rm 90}$ duration, namely using the time-integrated data. We have run \texttt{rmfit} with the following phenomenological models which are widely used: the Band model  \citep{Band1993_413}, SBPL model \citep{Ryde1999_39}, and power-law model with an exponential cutoff (CPL) \citep[see, e.g.,][]{Kaneko2006_166}. We have also studied spectral deviations from these models in the form of Band combined with PL \citep{Gonzalex2003_424, Abdo2009, Ackermann2010_716, Guiriec2010_725} or BB \citep{Guiriec2011_727, Guiriec2013_770, Guiriec2016_819, Guiriec2017_846, Axelsson2012_757}, and SBPL combined with BB \citep{Dirirsa2017, Ravasio2018_613}. The details of the functional forms of these models are described in Appendix~\ref{spectral_models}. Like the Cash-statistic, the C-stat is suitable for the analysis of counts that are Poisson distributed\footnote{https://heasarc.nasa.gov/docs/xanadu/xspec/xspec11/manual/\\node57.html}. It has been proposed to mimic a $\chi^2$ statistic and to provide a straightforward goodness of fit in the large sample limit. Assuming that the C-stat is $\chi^2$ distributed in the low count regime only provides an approximate judgement of the fit quality. Like the Cash-statistic, the C-stat can also be used to perform hypothesis testing between two nested models. Following the Wilks' theorem \citep{Wilks1938_9} and again assuming that the large sample limit is reached, we decide that a new spectral component (BB or PL, which both have two parameters) is required by the data in addition to the main component (Band, SBPL or CPL, chosen from the model with the lowest C-stat value) if they cause a decrease in C-stat that is larger than 25 (i.e. approximately $5\sigma$ for a $\chi^2$ with 2 degrees of freedom). We use the same criterion to compare the Band+BB and Band+BB+PL models, or the Band+CPL and Band+CPL+PL models. The spectral parameters obtained from the best models along with the C-stat values are presented in Table~\ref{tab:modelfit}. Often the brighter GRBs require more complex models such as SBPL\,+\,BB, Band\,+\,BB, Band\,+\,CPL\,+\,PL; etc., for fitting their spectra.  Fig.~\ref{fig:models} shows the $\nu F_\nu$ energy spectrum using the best model of spectral fit for each GRB in Table~\ref{tab:modelfit}. The top left panel shows the spectra fitted with Band or Band + BB and the top right panel shows the spectra fitted with SBPL or SBPL + BB.  The bottom left panel shows GRB\,160625B spectrum fitted with Band\,+\,BB\,+\,PL and GRB\,160509A spectrum fitted with Band\,+\,CPL\,+\,PL.  The bottom right panel shows the spectra of GRB\,130427A, GRB\,090926A, GRB\,090902B and  GRB\,090510, all fitted with Band\,+\,PL. The shaded regions correspond to the $1\sigma$ confidence intervals of the models.

\subsection{Isotropic energy calculation}
We computed the isotropic radiated energy $E_{\rm iso}$ in the source rest frame at a luminosity distance $d_{\rm L}$ as  
\begin{equation}
E_{\mathrm{iso}} = \frac{4\pi d_{\mathrm{L}}^2}{1+z}S_{\mathrm{bolo}},  
\label{Eiso_sbolo}
\end{equation}
where 
\begin{equation} 
S_{\mathrm{bolo}} = T_{90}\displaystyle\int_{{E_{\mathrm{min}}}/{(1+z)}}^{{E_{\mathrm{max}}}/{(1+z)}}EN_{\mathrm{i}}(E)\,dE
\label{sbolo}
\end{equation}
is the bolometric fluence integrated over the minimum photon energy $E_{\mathrm{min}}$ = 1 keV and the maximum photon energy $E_{\mathrm{max}} = 10^4$ keV or $10^5$ keV.  Here $N_{\mathrm{i}}(E)$ represents the  best-fit spectral model discussed previously.  The luminosity distance strongly depends on the cosmological model.  Assuming a flat $\Lambda$CDM cosmology with $\Omega_{\mathrm{m}} = 1 - \Omega_{\Lambda}$, the $d_{\mathrm{L}}$ can be expressed as 
\begin{equation}\label{eq_dL}
d_{\mathrm{L}} = (1+z)\displaystyle\frac{c}{H_{\mathrm{0}}}\int_0^z \frac{dz'}{\displaystyle\sqrt{(1-\Omega_{\Lambda})(1+z')^3+\Omega_{\Lambda}}}.  
\end{equation}

We have used Monte Carlo simulations to estimate the errors on $S_{\mathrm{bolo}}$ and $E_{\mathrm{iso}}$.  In particular, we assume the parameters of the spectral models follow a multivariate Gaussian function. Using the covariance matrix obtained from a spectral fit, we generate $10^4$ sets of random values for the parameters to calculate $S_{\mathrm{bolo}}$ in equation~(\ref{sbolo}). We select $68.27\%$ confidence intervals from the resulting distribution of $S_{\mathrm{bolo}}$ values to evaluate its error.  

The results of $S_{\mathrm{bolo}}$ and $E_{\mathrm{iso}}$ are reported in Table~\ref{tab:Sbolo}.   The values of $S_{\mathrm{bolo}}^*$(F10) and $E_{\mathrm{iso}}^*$(F10) correspond to the energy interval 1--$10^4$ keV, while the values of $S_{\mathrm{bolo}}^{*}$(F100) and $E_{\mathrm{iso}}^{*}$(F100) correspond to the energy interval 1--$10^5$ keV.  The intrinsic peak energies of the $\nu F_\nu$ spectra are reported as $E_{\mathrm{i,p}}$ in Table~\ref{tab:Sbolo}. In all cases, $E_{\mathrm{i,p}} = (1+z)E_\mathrm{p}$.

\begin{table*}[htb!]
\begin{center}
\caption{Intrinsic peak energy ${E_{\mathrm{i,p}}}$ and isotropic radiated energy ${E_{\mathrm{iso}}}$ for the \textit{Fermi} GRB sample.}
\scalebox{.71}{
\begin{tabular}{@{}l*{45}{l}}  
\hline \hline
GRB  name   & $z$  & $E_{\mathrm{i,p}}$  & $E_{\mathrm{iso}}^{*}$(F10)     & $S_{\mathrm{bolo}}^{*}$(F10)           &           & $E_{\mathrm{iso}}^{*}$(F100)  &$S_{\mathrm{bolo}}^{*}$(F100) &  & References for redshift     \\
            &      &  (keV)              &  (10$^{52}$ erg)           & (10$^{-5}$ erg/cm$^{2}$)   &$\mu \pm \sigma_{\mu}^{\dagger}$                        & (10$^{52}$ erg)  & (10$^{-5}$ erg/cm$^{2}$)     &    $\mu \pm \sigma_{\mu}^{\ddagger}$               \\
\hline
GRB\,170405A & 3.51          & 1424.42 $\pm$ 35.24   & 240.00 $\pm$ 2.41            & 9.24 $\pm$ 0.09      & 46.22	$\pm$ 1.27  & 293.00 $\pm$ 4.87            & 11.28 $\pm$ 0.19     &46.19	$\pm$ 1.20   &\cite{170405A_3.51_Postigo2017GCN_20990}       \\
GRB\,170214A & 2.53           & 2119.788 $\pm$ 119.06 & 338.00 $\pm$ 4.36            & 22.40 $\pm$ 0.29    &45.49  $\pm$ 1.30     & 425.00 $\pm$ 6.10             & 28.22 $\pm$ 0.40 &45.47	$\pm$ 1.21   &\cite{170214A_2.53_Kruehler2017GCN_20686}      \\
GRB\,160625B & 1.406         & 1546.86 $\pm$ 37.25   & 435.01 $\pm$ 6.06            & 83.54 $\pm$ 1.16     &43.25	$\pm$ 1.27   & 494.13 $\pm$ 7.34             & 94.87 $\pm$ 1.42   &43.31	$\pm$ 1.20     &\cite{160625B_1.406_Xu2016GCN_19600}       \\
GRB\,160509A & 1.17          & 19334.10 $\pm$ 652.25 & 182.68 $\pm$ 4.97            & 49.91 $\pm$ 1.36     &46.89	$\pm$ 1.67   & 364.81 $\pm$ 13.81             & 99.73 $\pm$ 3.79  &46.57	$\pm$ 1.26   &\cite{160509A_1.17_Tanvir2016GCN_19419}     \\
GRB\,150514A & 0.807         & 137.84 $\pm$ 14.93    & 1.26 $\pm$ 0.05              & 0.71 $\pm$ 0.03      &45.05	$\pm$ 1.45   & 1.37 $\pm$ 0.07               & 0.78 $\pm$ 0.04    &44.92	$\pm$ 1.23    &\cite{150514A_0.807_Postigo2015GCN_17822}       \\
GRB\,150403A & 2.06          & 2428.51 $\pm$ 160.80  & 85.20 $\pm$ 1.81             & 8.10 $\pm$ 0.17      & 46.61	$\pm$ 1.31  & 95.30 $\pm$ 2.72              & 9.06 $\pm$ 0.27     &46.73	$\pm$ 1.21    &\cite{150403A_2.06_Pugliese2015GCN_17672}       \\
GRB\,150314A & 1.758         & 985.66 $\pm$ 13.20    & 72.70 $\pm$ 0.96             & 9.20 $\pm$ 0.12      &45.22	$\pm$ 1.26    & 76.00 $\pm$ 1.72              & 9.61 $\pm$ 0.22   & 45.33	$\pm$ 1.20   &\cite{150314A_1.758_Postigo2015GCN_17583}      \\
GRB\,141028A & 2.33          & 1320.18 $\pm$ 50.90   & 64.00 $\pm$ 0.74             & 4.89 $\pm$ 0.06      & 46.48	$\pm$ 1.27   & 79.80 $\pm$ 1.24              & 6.10 $\pm$ 0.09    &46.42	$\pm$ 1.20    &\cite{141028A_2.33_Xu2014GCN_16983}     \\
GRB\,131231A & 0.6439        & 370.15 $\pm$ 4.97     & 19.20 $\pm$ 0.13             & 17.42 $\pm$ 0.12     &42.73	$\pm$ 1.31   & 20.10 $\pm$ 0.20              & 18.23 $\pm$ 0.18   &42.74	$\pm$ 1.21   &\cite{131231A_0.6439_Cucchiara2014GCN_15652}   \\
GRB\,131108A & 2.40          & 1163.20 $\pm$ 28.54   & 66.80 $\pm$ 0.65             & 4.85 $\pm$ 0.05      &46.35	$\pm$ 1.26     & 89.90 $\pm$ 1.19              & 6.53 $\pm$ 0.09  & 46.20	$\pm$ 1.20   &\cite{131108A_2.40_Postigo2013GCN_15470}       \\
GRB\,130518A & 2.49          & 1601.40 $\pm$ 32.19   & 167.00 $\pm$ 1.53            & 11.40 $\pm$ 0.11     &45.86	$\pm$ 1.27   & 189.00 $\pm$ 2.48             & 12.92 $\pm$ 0.17   & 45.92	$\pm$ 1.21   &\cite{130518A_2.49_Sanchez-Ramirez2013GCN_14685}  \\
GRB\,130427A & 0.3399        & 294.25 $\pm$ 5.86     & 9.29 $\pm$ 0.06             & 31.72 $\pm$ 0.20      &41.56	$\pm$ 1.33   & 10.65 $\pm$ 0.12         & 36.34 $\pm$ 0.39   &41.46	$\pm$ 1.21    &\cite{130427A_0.34_Levan2013GCN_14455}       \\
GRB\,120624B & 2.2           & 1214.47 $\pm$ 26.24   & 242.00 $\pm$ 2.95            & 20.49 $\pm$ 0.25     &44.78	$\pm$ 1.26    & 267.00 $\pm$ 4.65             & 22.63 $\pm$ 0.39   &44.85	$\pm$ 1.20    &\cite{120624B_2.2_Postigo2013_557}    \\
GRB\,110721A & 3.512         & 8675.78 $\pm$ 852.66  & 160.0 $\pm$ 2.33            & 6.14 $\pm$ 0.09       &48.94	$\pm$ 1.50     & 243.00 $\pm$ 9.35             & 9.35 $\pm$ 0.34   & 48.85	$\pm$ 1.24   &\cite{110721A_3.512_Berger2011GCN_12193}       \\
GRB\,100728A & 1.567         & 797.62 $\pm$ 18.05    & 75.00 $\pm$ 1.06             & 11.74 $\pm$ 0.17     &44.61	$\pm$ 1.26    & 82.50 $\pm$ 1.74              & 12.92 $\pm$ 0.27   &44.64	$\pm$ 1.20   &\cite{100728A_1.567_Kruehler2010GCN_14500}    \\
GRB\,100414A & 1.368         & 1370.82 $\pm$ 27.68   & 58.70 $\pm$ 0.77             & 11.88 $\pm$ 0.16     &45.19	$\pm$ 1.27   & 63.50 $\pm$ 1.24              & 12.86 $\pm$ 0.25    & 45.30	$\pm$ 1.20  &\cite{100414A_1.368_Cucchiara2010GCN_10608}      \\
GRB\,091208B & 1.063         & 202.63 $\pm$ 20.10    & 2.26 $\pm$ 0.12              & 0.75 $\pm$ 0.04      &45.63	$\pm$ 1.39    & 2.37 $\pm$ 0.17               & 0.78 $\pm$ 0.06    & 45.59	$\pm$ 1.23 &\cite{091208B_1.063_Wiersema2009GCN_10263}     \\
GRB\,091127  & 0.49          & 60.32 $\pm$ 1.93      & 1.41 $\pm$ 0.02              & 2.25 $\pm$ 0.04      &42.55	$\pm$ 1.61    & 1.52 $\pm$ 0.03               & 2.42 $\pm$ 0.05    & 42.36	$\pm$ 1.25  &\cite{091127_0.49_Cucchiara2009GCN_10202}        \\
GRB\,091003A & 0.8969        & 857.81 $\pm$ 33.08    & 9.58 $\pm$ 0.16              & 4.43 $\pm$ 0.08      &45.43	$\pm$ 1.26    & 10.20 $\pm$ 0.21              & 4.70 $\pm$ 0.10    &45.51	$\pm$ 1.20   &\cite{091003_0.8969_Cucchiara2009GCN_10031}       \\
GRB\,090926A & 2.1062        & 868.63 $\pm$ 13.85     & 196.00 $\pm$ 1.39            & 17.90 $\pm$ 0.13    &44.47	$\pm$ 1.26    & 246.00 $\pm$ 3.26             & 22.43 $\pm$ 0.30   & 44.37	$\pm$ 1.20  &\cite{090926A_2.1062_Malesani2009GCN_9942}      \\
GRB\,090902B & 1.822         & 2146.57 $\pm$  21.71  & 329.00 $\pm$ 1.87            & 39.05 $\pm$ 0.22     & 44.66	$\pm$ 1.29    & 349.00 $\pm$ 3.35             & 41.45 $\pm$ 0.40   &44.83	$\pm$ 1.21    &\cite{090902B_1.822_Cucchiara2009GCN_9873}        \\
\textbf{GRB\,090510}  & \textbf{0.903}         & \textbf{7227.15 $\pm$ 731.88}  & \textbf{4.15 $\pm$ 0.18}              & \textbf{1.89 $\pm$ 0.08}       & -    & \textbf{7.19 $\pm$ 0.34}               & \textbf{3.28 $\pm$ 0.16}         &  -&\cite{GRB090510_0.903_Rau2009GCN_9353}  \\
GRB\,090424  & 0.544          & 236.91 $\pm$ 4.55     & 4.45 $\pm$ 0.07              & 5.72 $\pm$ 0.09        &43.30	$\pm$ 1.36   & 4.77 $\pm$ 0.12               & 6.13 $\pm$ 0.15     &43.25	$\pm$ 1.21   &\cite{090424_0.544_Chornock2009GCN_9243}     \\
GRB\,090328  & 0.736         & 1221.71 $\pm$ 81.87   & 11.60 $\pm$ 0.29             & 7.99 $\pm$ 0.20         &45.14	$\pm$ 1.27  & 14.20 $\pm$ 0.45              & 9.82 $\pm$ 0.31      &45.10	$\pm$ 1.21  &\cite{090328_0.736_Cenko2009GCN_9053}     \\
GRB\,090323  & 3.57          & 2060.09 $\pm$ 138.07  & 430.00 $\pm$ 10.40           & 15.76 $\pm$ 0.39        &46.12	$\pm$ 1.29   & 535.00 $\pm$ 17.20            & 19.64 $\pm$ 0.62    &46.10	$\pm$ 1.21   &\cite{090323_3.57_Chornock2009GCN_9028}       \\
GRB\,080916C & 4.35$\pm$0.15  & 6953.87 $\pm$ 1188.77  & 380.00 $\pm$ 8.61           & 10.40 $\pm$ 0.24       & 48.28	$\pm$ 1.47    & 605.00 $\pm$ 24.80            & 16.54 $\pm$ 0.68   & 48.11	$\pm$ 1.25  & \cite{Greiner2009_498}            \\\\ 
\hline
\label{tab:Sbolo}
\end{tabular}}
\end{center}

\scriptsize{\textbf{Notes}. The bolometric fluence $S_{\rm bolo}^{*}$(F10) and isotropic energy $E_{\rm iso}^{*}$(F10) are computed for the energy range 1-$10^4$ keV using equations (\ref{sbolo}) and (\ref{Eiso_sbolo}).  $S_{\rm bolo}^{*}$(F100) and $E_{\rm iso}^{*}$(F100) are computed for the energy range 1-$10^5$ keV.  ${E_{\mathrm{i,p}}} = (1+z)E_{\rm 0}$ for the SBPL spectral fits, with e-folding energy $E_{\rm 0}$.   ${E_{\mathrm{i,p}}} = (1+z)E_{\rm p}$ for the Band or CPL spectral fits, with peak energy $E_{\rm p}$. \\\\
       
      }
\label{redshift}
\end{table*}

\begin{figure}[htb]
\begin{center}
\includegraphics[width=8.5 cm]{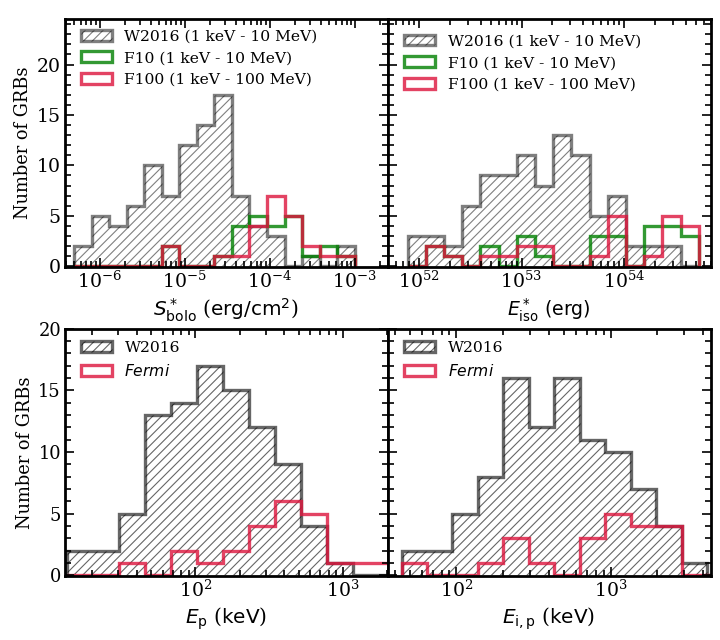}
\caption{{\it Top panels --} Distributions of bolometric fluence and isotropic energy of ${\it Fermi}$ ($S_{\rm bolo}^*$(F10) and $E_{\mathrm{iso}}^*$(F10) computed in the 1 keV--10 MeV energy range and F100, the same computed in the 1 keV--100 MeV energy range) and of W2016 samples.  {\it Bottom panels --}  Distributions of the observed peak energy (left) and of intrinsic peak energy (right) for the ${\it Fermi}$ and W2016 bursts.}
\label{EisoSboloHistogram}
\end{center}
\end{figure}

Figure~\ref{EisoSboloHistogram} shows the distributions of bolometric fluence, isotropic radiated energy and peak energy of the $\nu F_\nu$ spectra of {\it Fermi} GRBs and those in the W2016 sample.  There are small differences between the bolometric fluence $S_{\rm bolo}^{*}$(F10) and $S_{\rm bolo}^{*}$(F100) for the {\it Fermi} sample (green and red histograms, respectively, in the top-left panel), but both distributions peak at a higher fluence than the W2016 sample (grey histogram). Note that the W2016 sample is our reanalysis of data from \cite{Wang2016_585} as explained in Appendix~\ref{W26_updated}. Therefore, it forms a part of our work. The difference between F10 and F100 is due to high-energy emission in the {\it Fermi}-LAT range, often requiring additional spectral component(s). On the other hand, the difference between the {\it Fermi} and W2016 samples indicates that the GRBs detected by {\it Fermi}-LAT are more fluent. Similar conclusions can be drawn for the isotropic energy distributions (top-right panel), although the difference between the {\it Fermi} and W2016 samples is less dramatic. The mean values of the $E_\mathrm{p}$ and $E_\mathrm{i,p}$ distributions (bottom panels) are also higher in the {\it Fermi} sample than in the W2016 sample. 

\section{Amati relation between $E_{\mathrm{iso}}$ and $E_{\mathrm{i,p}}$}\label{sec_4}
\subsection{Fitted model}
The phenomenological Amati relation~\citep{Amati2002_81A} between $E_{\mathrm{iso}}$ and $E_{\mathrm{i,p}}$ is of the form
\begin{equation}
E_{\rm iso} \propto \left( \frac{E_{\mathrm{i,p}}}{E_0} \right)^m E_{0,\rm iso}
\label{Amati_nonlinear}
\end{equation}
where $m$ is the power law index, $E_{\rm 0}$ and $E_{0,\rm iso}$ are reference energies. Following \citet{Wang2016_585} and \citet{Demianski2017_693} we use a linearized Amati relation 
\begin{eqnarray}
y = mx + k \,\,;\, y \equiv \log_{10}\frac{E_{\mathrm{iso}}}{E_{0,\rm iso}}\,,\, x \equiv \log_{10}\frac{E_{\mathrm{i,p}}}{E_{\rm 0}}\,.
\label{Amati_linear}
\end{eqnarray}
We use only LGRBs for Amati relations fits. A preliminary fit to the $E_{\mathrm{iso}}^*$(F10) and $E_{\mathrm{i,p}}$ data for the {\it Fermi} sample in Table~\ref{tab:Sbolo} with $E_{0,\rm iso} = 10^{52}$~erg and $E_{\rm 0} = 100$~keV allows us to calculate the so-called ``de-correlation''\footnote{https://fermi.gsfc.nasa.gov/ssc/data/analysis/scitools/python\\$\_$tutorial.html} value of $x$, at which the error on $y$ is the smallest. This value can be obtained from a simple error propagation in equation \eqref{Amati_linear} as $x_{\rm dec} = -C_{\rm km}/\sigma_{\rm m}^2$, where $C_{\rm km}$ is the co-variance of the parameters $k$ and $m$, and $\sigma_{\rm m}$ is the error on $m$. Setting $E_0$ at the corresponding de-correlation energy, $E_{\rm 0,dec} = 10^{x_{\rm dec}}100$~keV also removes the correlation between the parameters $m$ and $k$ (i.e. $C_{\rm km}\sim$0), which will allow us to discuss these parameters independently. The value of $E_{\rm 0,dec}$ for $E_{\rm iso}$ (F10 or F100) and $E_{\rm i,p}$ data is 950 keV. For the W2016 sample in Table~\ref{my-label}, however, this energy is 450~keV. We also calculate the de-correlation energy for the combined {\it Fermi} and W2016 data sets. Furthermore, the values of $E_{0,\rm dec}$ for the analysis of high redshift samples of GRBs listed in Table \ref{tab:fitlkhood_2}.  

\subsection{Likelihood analysis}
We have performed a likelihood analysis to extract not only the best-fit values of the parameters $m$ and $k$ in equation~(\ref{Amati_linear}) but also the extrinsic uncertainty $\sigma_{\rm ext}$ on $y$, which is treated as an unknown parameter. This may account for hidden parameters related to the physical origin of the Amati relation. We have taken the mean errors on $E_{\rm iso}$ and $E_{\rm i,p}$ and used $\log(E_{\rm iso} \pm \sigma_{E_{\rm iso}})$ and $\log(E_{\rm i,p} \pm \sigma_{E_{\rm i,p}})$ to get the errors on $x$ and $y$. This is also the procedure followed in \cite{Wang2016_585} and \cite{Demianski2017_693}. After propagating these errors, we get asymmetric errors in $m$ and $k$ of the Amati relation, which we have symmetrized by taking the mean. We have checked that the mean error and asymmetric errors are similar. Following \citet{DAgostini2005_11182D} we apply the log likelihood function $-\ln \mathcal{L}(m,k,\sigma_{\mathrm{ext}}) = L(m,k,\sigma_{\mathrm{ext}})$ to fit the $x=\log_{10} (E_{\rm i,p}/E_{0,\rm dec})$ and $y=\log_{10}(E_{\rm iso}/E_{0,\rm iso})$ data with equation~(\ref{Amati_linear}).  The functional form is given by
\begin{equation}
\begin{aligned}
{L}(m,k,\sigma_{\mathrm{ext}}) = {} & \frac{1}{2}\sum_{{i}}^N \ln (\sigma^2_{\mathrm{ext}}+\sigma^2_{y_{\mathrm{i}}}+m^2\sigma^2_{x_{\mathrm{i}}}) \\
      & + \frac{1}{2} \sum_{{i}}^N{\frac{(y_{\mathrm{i}}-mx_{\mathrm{i}}-k)^2}{(\sigma^2_{\mathrm{ext}}+\sigma^2_{\mathrm{y_{i}}}+m^2\sigma^2_{x_{\mathrm{i}}})}}\,,
      \end{aligned}
\label{DAgostini2005}
\end{equation}
where $\sigma_{\rm x_i}$ and $\sigma_{\rm y_i}$ are errors on the $x$ and $y$ data, respectively. We minimize this function to find the best-fit values of the parameters $m$, $k$ and $\sigma_{\mathrm{ext}}$. These are listed in Table~\ref{tab:fitlkhood_2} for the {\it Fermi} and W2016 samples, and for the combination of the two samples.  

\begin{figure}[htb]
\includegraphics[width=8.5 cm]{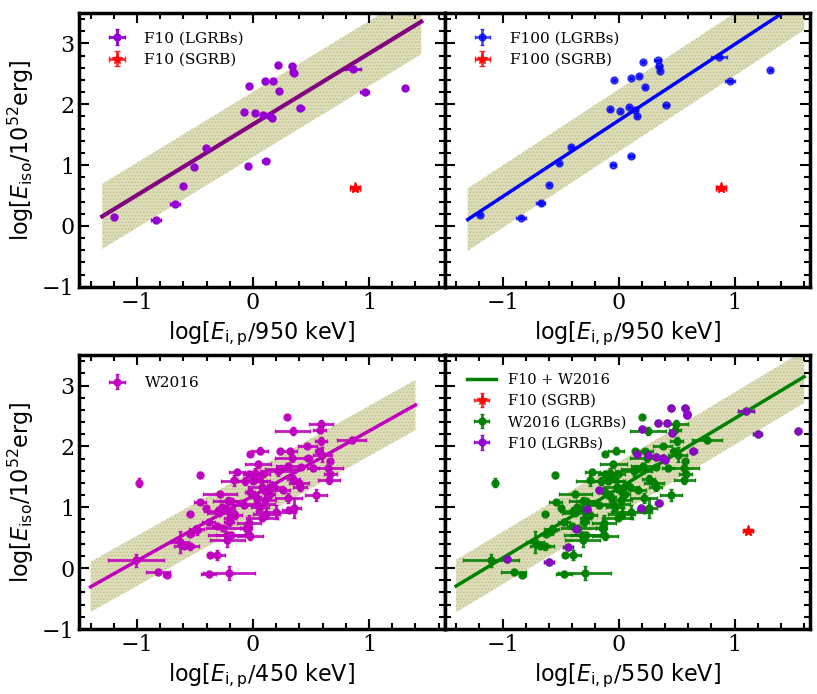}
\caption{Amati relation fits (solid lines) to different data samples in the $\log {E_{\rm iso}}-\log {E_{\rm i,p}}$ plane. The shaded regions correspond to one-sigma scatter.
\textit{Top left panel --} \textit{Fermi} 25 LGRBs with ${E_{\rm iso}}$ computed in the 1 keV--10 MeV energy range (F10). \textit{Top right panel --} \textit{Fermi} 25 LGRBs with ${E_{\rm iso}}$ computed in the 1 keV--100 MeV energy range (F100). \textit{Bottom left panel --} 94 LGRBs in the W2016 sample. \textit{Bottom right panel --} Joint fit to the {\it Fermi} F10 and W2016 data reported in Table  \ref{tab:fitlkhood_2}.}
\label{Cosmolo}
\end{figure}

\begin{figure}[hbt!]
\begin{center}
\includegraphics[width=8.10 cm]{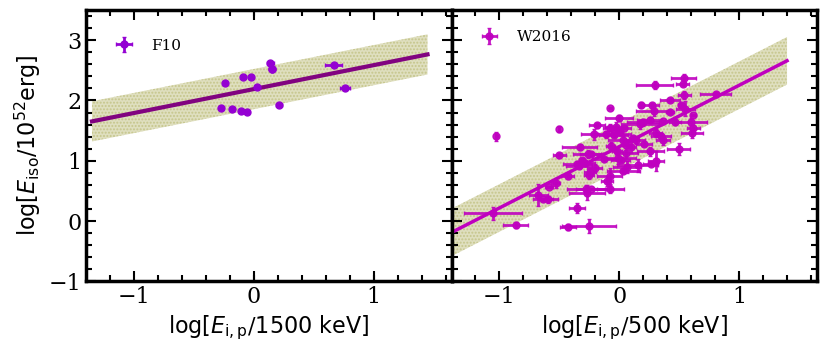}
\includegraphics[width=4.30 cm]{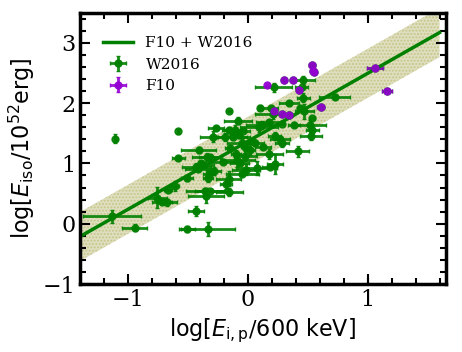}
\caption{Amati relation fits (solid lines) to different data samples with redshift $>1.414$ in the $\log {E_{\rm iso}}-\log{E_{\rm i,p}}$ plane. The sample descriptions are the same as in Fig.~\ref{Cosmolo}.} 
\label{fig:Cosmolo_1}
\end{center}
\end{figure}

To determine the uncertainties of a fit parameter $q_{\mathrm{i}}$, as in \citet{Demianski2011_415}, we evaluate the marginalized likelihood function $\mathcal{L_{\mathrm{i}}}(q_{\mathrm{i}})$ by integrating over the other parameters.  Then the median value for the parameter $q_{\mathrm{i,med}}$ is found from the integral
\begin{equation}
\int_{q_{\mathrm{i,min}}}^{q_{\mathrm{i,med}}}  \mathcal{L_{\mathrm{i}}}(q_{\mathrm{i}})dq_{\mathrm{i}} = \frac{1}{2}\int_{q_{\mathrm{i,min}}}^{q_{\mathrm{i,max}}} 
\mathcal{L_{\mathrm{i}}}(q_{\mathrm{i}})dq_{\mathrm{i}} \,, 
\end{equation}     
where $q_{\mathrm{i,min}}$ and $q_{\mathrm{i,max}}$ are the minimum and maximum value of the parameter, respectively.  The 1~$\sigma$ or 68.27$\%$ confidence interval ($q_\mathrm{i,l}, q_\mathrm{i,h}$) of the parameters are then found by solving the integral \citep{DAgostini2005_11182D}
\begin{equation}
\displaystyle\int_{q_{\mathrm{i,l}}}^{q_{\mathrm{i,med}}}  \mathcal{L_{\mathrm{i}}}(q_{\mathrm{i}})dq_{\mathrm{i}} = \displaystyle\frac{1}{2} (1-\eta) \int_{q_{\mathrm{i,min}}}^{q_{\mathrm{i,max}}}  \mathcal{L_{\mathrm{i}}}(q_{\mathrm{i}})dq_{\mathrm{i}} \,, 
\end{equation}
\begin{equation}
\displaystyle\int_{q_{\mathrm{i,med}}}^{q_{\mathrm{i,h}}}  \mathcal{L_{\mathrm{i}}}(q_{\mathrm{i}})dq_{\mathrm{i}}= \displaystyle\frac{1}{2} (1-\eta) \displaystyle\int_{q_{\mathrm{i,min}}}^{q_{\mathrm{i,max}}}  \mathcal{L_{\mathrm{i}}}(q_{\mathrm{i}})dq_{\mathrm{i}} \,, 
\end{equation}
where $\eta$ = 0.6827. Finally, we have calculated the mean of the upper and lower uncertainties for each parameter. Figure~\ref{Cosmolo} shows the Amati relation plotted against the ${\it Fermi}$ and W2016 data samples. As for comparisons of our fit parameters with those by other recent studies, \citet{Wang2016_585} found  $m = 1.48\pm 0.09$ and $\sigma_{\rm ext} = 0.34\pm 0.01$ from 151 GRBs.  
%
\begin{table}[hbt!]
	\centering
	\caption{The best-fit parameters of the Amati relation fits to the full samples of GRBs  and  GRBs with redshift $z > 1.414$. $\rho$ is the partial correlation coefficient.}  
	\scalebox{0.74}{
		\begin{tabular}{@{}l*{45}{l}}  
			\hline \hline 
			Full LGRB  & No.~of  & $\rho$ & $E_{0,\rm dec}$ &  &  &  \\ 
			samples    & GRBs    &      & (keV) & $m$ & $k$ & ${\sigma_{\rm ext}}$ \\ \hline
			F10        & 25      & 0.65 & 950   & 1.16 $\pm$ 0.37 & 1.67 $\pm$ 0.16  &  0.47 $\pm$ 0.12  \\
			F100       & 25      & 0.70 & 950	& 1.25 $\pm$ 0.33 & 1.73 $\pm$ 0.18  & 0.45 $\pm$ 0.13  \\
			W2016      & 94      & 0.71 & 450	& 1.07 $\pm$ 0.20 & 1.19 $\pm$ 0.08 & 0.38 $\pm$ 0.06  \\
			F10+W2016  & 119     & 0.77 & 550	& 1.15 $\pm$ 0.16 & 1.31 $\pm$ 0.07 & 0.41 $\pm$  0.05 \\ \hline  
			$z>1.414$ samples \\\hline
			F10        & 14      & 0.10 & 1500 & 0.40 $\pm$ 0.63 & 2.19 $\pm$ 0.13  &  0.26  $\pm$ 0.09   \\
			W2016      & 84      & 0.71 & 500	&  1.02 $\pm$ 0.17 & 1.22 $\pm$ 0.07 & 0.37 $\pm$  0.05  \\
			F10+W2016  & 98      & 0.77 & 550	 & 1.13 $\pm$ 0.17  & 1.37 $\pm$ 0.08 & 0.38 $\pm$  0.05
			\\ \hline          
	\end{tabular}}
	\label{tab:fitlkhood_2}
\end{table}
Similarly, \citet{Demianski2011_415} found $m = 1.52$ and $\sigma_{\rm ext} = 0.41$ by analyzing 109 GRBs. The uncertainty on the parameter $y$ is estimated as \citep{Wang2016_585,Demianski2017_693} 
\begin{equation}\label{sigma_Eiso}
\sigma_{\rm y} = \sqrt{\sigma_{k}^2+m^2\sigma_{x}^2+\sigma_{m}^2x^2+\sigma_{\mathrm{ext}}^2} \,,
\end{equation} 
where $x=\log_{10}(E_{\rm i,p}/E_{0,\rm dec}$).  

The results of our fit to the linearized Amati relation are shown in Figs.~\ref{Cosmolo} and \ref{fig:Cosmolo_1}, and are listed in Table~\ref{tab:fitlkhood_2}. The shaded region in Figs.~\ref{Cosmolo} and \ref{fig:Cosmolo_1} shows the $\pm 1\sigma_{\rm y}$ uncertainties on the Amati relation. It should be noted that this error is fully dominated by the extrinsic term $\sigma_{\rm ext}$ (see equation \ref{sigma_Eiso}).  We have fitted data from the 25 \textit{Fermi} LGRBs (F10 and F100 samples in the top left and right panels of Fig.~\ref{Cosmolo}, respectively), the W2016 sample of 94 GRBs (bottom left panel of Fig.~\ref{Cosmolo}) and a combination of the F10 and W2016 samples (bottom right panel of Fig.~\ref{Cosmolo}). Note that we expect the parameter $k$ to be different for different samples, due to a difference in the de-correlation energy. The relevant parameters to be compared among different samples are the slope $m$ and scatter $\sigma_{\rm ext}$, which are within errors for the F10, F100, W2016 and F10+W2016 samples. Table~\ref{tab:fitlkhood_2} also lists the de-correlation energy $E_{0,\rm dec}$ for each sample and their combination as well as partial correlation coefficient between ${E_{\mathrm{i,p}}}$ and ${E_{\mathrm{iso}}}$.  Analysis for the full data set provides a partial correlation coefficient ($\rho$) of 0.65 and 0.70 for samples F10 and F100, respectively, which are highly significant. 

We have also studied the Amati relation for GRBs with redshift $z > 1.414$, i.e., beyond the measured redshift of the supernovae data sample SNe U2.1 used for cosmology~\citep{Suzuki2012_746}.  This redshift cut leaves 14 {\it Fermi} LGRBs and 85 W2016 LGRBs for analysis.  The results of the fits are shown in Fig.~\ref{fig:Cosmolo_1} and are listed in Table~\ref{tab:fitlkhood_2}.  The reduced F10 sample itself does not provide significant constraints on the fit parameter $m$ and the corresponding partial correlation coefficient is also not large. On the other hand, the combination of the reduced F10 and W2016 samples gives the parameter values similar to those obtained from fitting all LGRBs. The partial correlation coefficient is also very high in this case. The reduced W2016 sample itself gives similar results as the full W2016 sample.

Our results show that the Amati relation holds for the bright GRBs detected by \textit{Fermi}, which provides the best energy coverage ever for the study of GRB prompt emission. This is an important confirmation since the W2016 and F10 samples are different in terms of energy coverage and redshift ranges. The W2016 sample is the sum of different sets of observed GRBs and it contains $\sim$4 times more GRBs than the F10 sample. This turns into two times smaller errors on the Amati parameters. While the W2016+F10 sample is thus dominated by the W2016 sample, adding the F10 sample still increases the accuracy on the Amati relation parameters and the correlation coefficient as seen in Table~\ref{tab:fitlkhood_2}. 
 
\section{The Hubble diagram and constraints on cosmological parameters}\label{sec_5}
\subsection{Analysis procedure and results}

\begin{figure*}[htb!]
\centering
\includegraphics[height=7.0 cm]{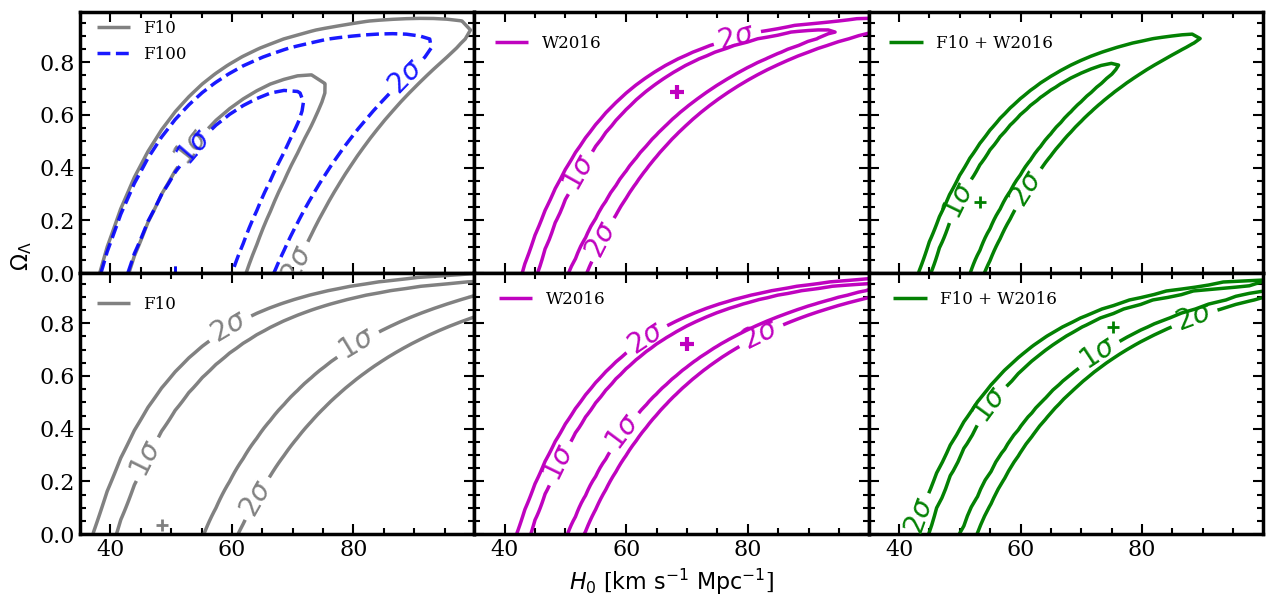}
\caption{Constraints on the cosmological parameters $H_0$ and $\Omega_{\Lambda}$ from the Amati relation for different samples of LGRBs.  The top and bottom panels correspond to the samples with all redshifts and with redshifts $> 1.414$, respectively.  The $1 \sigma$ and $2\sigma$ confidence level contours are determined by following $\Delta\chi^2 \equiv \chi^2 - \chi^2_{\rm min} \leq 2.30$ and $6.18$, respectively. The plus sign indicates the best fit location. }
\vskip 0.25 cm
\label{fig:Cosmo_par1}
\end{figure*}

Once the parameters are obtained by fitting the linearized Amati relation (Section \S\ref{sec_4}), we can use the LGRBs as cosmological probes. In particular we can invert the relation in equation~(\ref{Eiso_sbolo}) to obtain the luminosity distance as 
\begin{equation}
d_{\rm L} = \left[ \frac{1+z}{4\pi} \frac{E_{0,\rm iso}}{S_{\rm bolo}} 10^k 
\left( \frac{E_{\mathrm{i,p}}}{E_{0,\rm dec}} \right)^m  \right]^{1/2} \,.
\label{dL_Amati}
\end{equation}
We then use this $d_L$ for each GRB to construct the GRB Hubble diagram, i.e., the distance modulus as a function of the redshift, $\mu(z) = 5\log(d_{\rm L}/{\rm 1\,Mpc}) + 25$. The uncertainty on $\mu$ is given by
\begin{equation}
\sigma_{\mathrm{\mu}}(z) = \left[\left( \frac{5}{2}\sigma_{\log E_{\mathrm{iso}}} \right)^2+\left(\frac{5}{2\ln10}\frac{\sigma_{\mathrm{S_{bolo}}}}{S_{\rm bolo}}\right)^2\right]^{1/2} \,.
\label{sigma_modulus}
\end{equation}
Here $\sigma_{\log E_{\rm iso}}$ is the propagated uncertainties on $E_{\rm iso}$ computed from equations \eqref{Amati_linear} and \eqref{sigma_Eiso}, which is given by 
\begin{eqnarray}
\sigma^2_{\log E_{\mathrm{iso}}} &=& \left(\sigma_{\rm{m}}\log\frac{E_{\mathrm{i,p}}}{E_{0,\rm dec}}\right)^2+\left(\frac{m}{\ln 10} \frac{\sigma_{E_{\rm{i,p}}}}{E_{\mathrm{i,p}}}\right)^2 \nonumber \\
&& + \sigma_{\rm{k}}^2 + \sigma^2_{\mathrm{ext}}.
\end{eqnarray}

We constrain the parameters of a flat $\Lambda$CDM cosmological model using GRB data. In particular, we use the Amati relation parameters from Table~\ref{tab:fitlkhood_2} in equations~(\ref{sigma_modulus}) to calculate the uncertainty of the distance modulus. The best-fit density parameter $\Omega_{\Lambda}$ and Hubble parameter $H_0$ are  estimated by the minimization of the $\chi^2$ expression given by  
\begin{equation}\label{eq_lk_2}
\chi^2(H_0,\Omega_{\Lambda}) = \displaystyle\sum\limits_{i=0}^{N}\frac{\left(\mu{(z_{\rm i})}-\mu^{\mathrm{pred}}{(z_{\rm i}; H_0,\Omega_{\Lambda})}\right)^2}{\sigma^2_{\mu_{(z_{\rm i})}}}  \,.
\end{equation}
Here $z_{\rm i}$ is the measured redshift for each GRB, $\mu{(z_{\rm i})}$ is the distance modulus, $\sigma_{\mu_{\rm zi}}$ is the uncertainty of the observed distance modulus obtained from equation \eqref{sigma_modulus} and $\mu^{\rm pred}{(z_{\rm i},H_0,\Omega_{\Lambda})} = 5 \log (d_L(H_0,\Omega_{\Lambda})/{\rm{1\,Mpc}}) + 25$ is a theoretically predicted value of the distance modulus computed from equation \eqref{eq_dL}.

The results from our analysis are listed in Table~\ref{tab:cosmological_parameters}, giving the best-fit values of $H_0$ and $\Omega_\Lambda$ when it is possible to constrain those parameters. Fig.~\ref{fig:Cosmo_par1} shows $1\sigma$ and $2\sigma$ contours in the $H_0 -\Omega_\Lambda$ plane for different sample selections. We note that F10 and F100 samples by themselves cannot constrain $\Omega_\Lambda$ (Fig.~\ref{fig:Cosmo_par1} top left panel). 
The size of the W2016 sample is larger and corresponds to a wider range in redshift (see the top panel of Fig.~\ref{tab:histogram}). Since the $\Omega_\Lambda$ parameter determines the evolution of the luminosity distance with redshift (see equation \ref{eq_dL}), this sample provides better results and best-fit values of $H_0$ and $\Omega_\Lambda$ (Fig.~\ref{fig:Cosmo_par1} top middle panel) which are consistent with the current best-fit values from {\it Planck}~\citep{Planck2018} within errors. The combined F10+W2016 sample moves the best-fit point in the $H_0$--$\Omega_\Lambda$ plane (Fig.~\ref{fig:Cosmo_par1} top right panel) away from the W2016 sample alone, but still consistent within $1 \sigma$ error. 

The bottom panels of Fig.~\ref{fig:Cosmo_par1} shows the $1\sigma$ and $2\sigma$ contours in the $H_0 -\Omega_\Lambda$ plane for the same samples as in the top panels but only for $z > 1.414$. Note that the $1\sigma$ confidence intervals for F10+W2016 include {\it Planck} values. For instance, $H_0 = 75^{+24}_{-31}$ (W2016+F10) is fully compatible with $H_0 = 67.4 \pm 0.5$ ({\it Planck}) within errors. Similarly $\Omega_\Lambda = 0.78^{+0.17}_{-0.78}$ (W2016+F10) is also fully compatible with $\Omega_\Lambda$ = 0.714  within errors. Indeed, the best-fit values of these parameters for the W2016 sample alone are closer to the {\it Planck} values. A bias could be introduced by the F10 sample because the $\Omega_\Lambda$ value is not constrained at all due to the limited range in redshift of the {\it Fermi} sample. The F10 sample does not provide any sufficient constraint. 

\begin{table}[htb!]
 \caption{Constraints on ($H_0$, $\Omega_{\Lambda}$) for a flat Universe from the Amati relation, and reduced $\chi^2$/dof of the fits.}  
\centering
\scalebox{0.80}{
 \begin{tabular}{@{}l*{45}{l}} 
\hline\hline 
Reference (all sample)  		 	& $H_0$ [km s$^{-1}$ Mpc$^{-1}$] 		
& $\Omega_{\Lambda}$   &  $\chi^2/dof$     			\\ \hline
F10                    &   51$^{+25}_{-8}$     &  -   & -  \\
F100                   &  48$^{+21}_{-8}$     &  -  &  -    \\
W2016                  & 68$^{+26}_{-23}$     &  0.69$^{+0.2}_{-0.7}$     &      87.2/92                                   \\
F10 + W2016            & 53$^{+23}_{-8}$       &  0.27$^{+0.53}_{-0.27}$  &   111.0/117    \\
SNe U2.1         & 70$^{+0.6}_{-0.5}$    &   0.72 $\pm$ 0.03  & 562.3/578       \\ 
F10 + SNe U2.1         & 70$^{+0.6}_{-0.5}$     &   0.72 $\pm$ 0.03  & 580.4/603       \\
F10 + W2016 + SNe U2.1 & 70$\pm$ 0.5     &   0.72 $\pm$ 0.03 & 667.6/697	  \\ \hline 
Sample with $z>1.414$\\ \hline 
F10                     & 48$^{+51}_{-8}$      &   0.03$^{+0.95}_{-0.03}$    &  9.1/12  \\
 W2016                  & 70$^{+30}_{-26}$      &  0.72$^{+0.2}_{-0.7}$      &      78.4/82                                   \\ 
 F10 + W2016            & 75$^{+24}_{-31}$       &  0.78$^{+0.17}_{-0.78}$  &   89.4/96    \\
 F10 + SNe U2.1         & 70$^{+0.5}_{-0.6}$     &   0.72 $\pm$ 0.03   & 578.2/592       \\
 F10 + W2016 + SNe U2.1 & 70 $\pm$ 0.5     &   0.72 $\pm$ 0.03	& 648.8/676\\  \hline 
\end{tabular}}
 \label{tab:cosmological_parameters}
\end{table}

\begin{figure*}[htb!]
\begin{center}
\includegraphics[width=5.8 cm,height=5.cm,keepaspectratio]{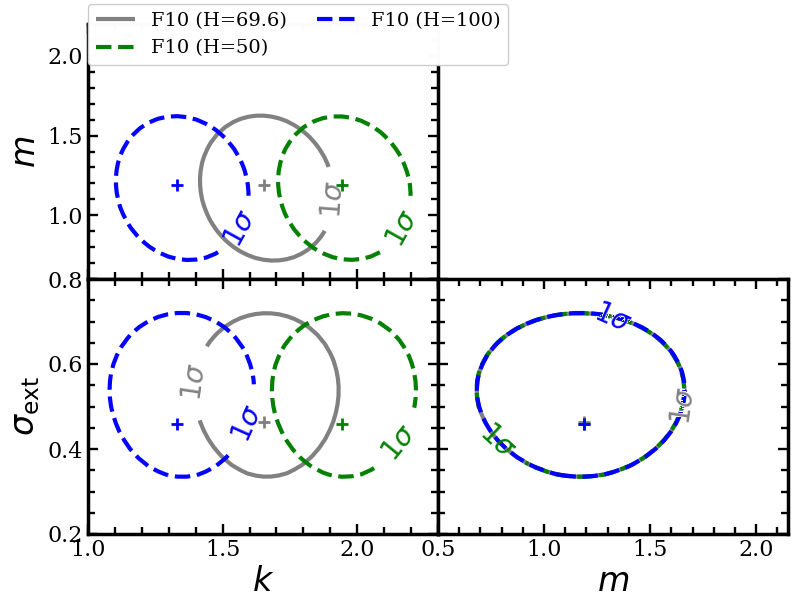}
\includegraphics[width=5.8 cm,height=5.cm,keepaspectratio]{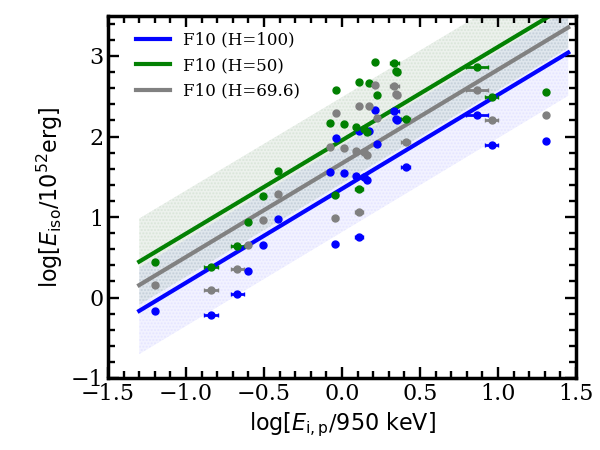}
\includegraphics[width=5.8 cm,height=5.cm,keepaspectratio]{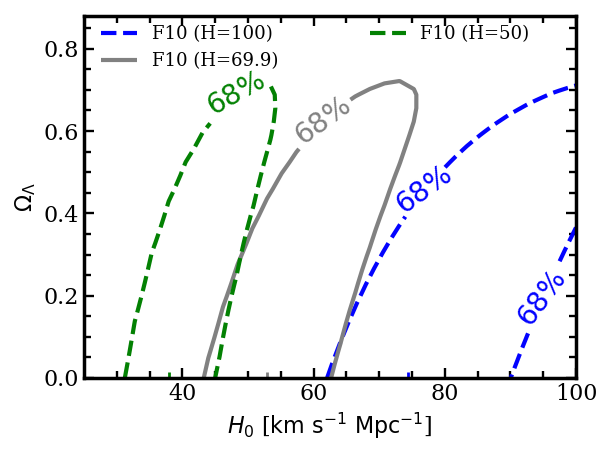}
\caption{Results from the sensitivity study of the Amati relation and cosmological parameters on the initial choice of $H_0$ and $\Omega_\Lambda$ to calculate $E_{\rm iso}$. The grey lines correspond to initial $H_0 = 69.6$~km~s$^{-1}$~Mpc$^{-1}$ (default case).  The blue and green lines correspond to initial values of $H_0 = 100$ and $50$~km~s$^{-1}$~Mpc$^{-1}$, respectively.  The initial value of $\Omega_{\Lambda} = 0.714$ (default case) is the same for all cases.}
\label{fig:hubble_H0}
\end{center}
\end{figure*}
%
\begin{figure*}[htb!]
\begin{center}
\includegraphics[width=5.8 cm,height=5.cm,keepaspectratio]{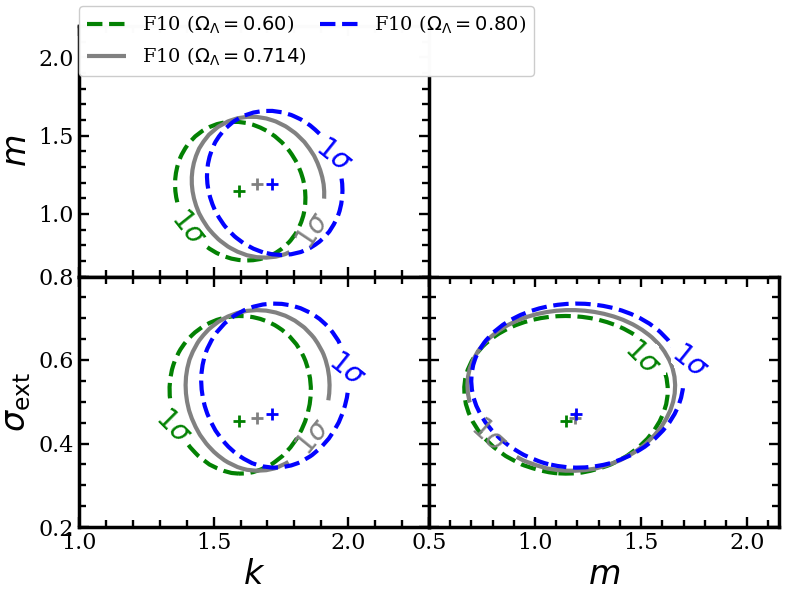}
\includegraphics[width=5.8 cm,height=5.cm,keepaspectratio]{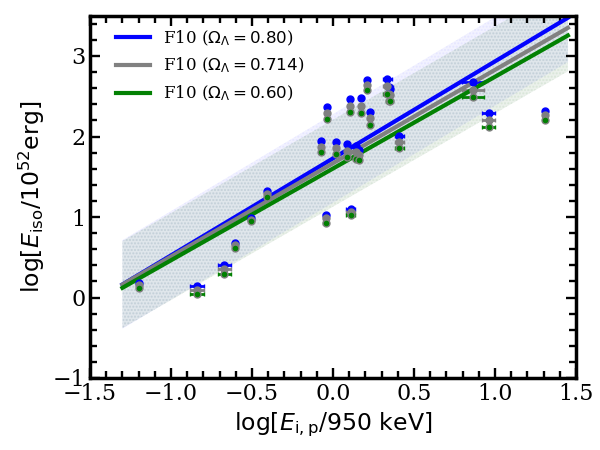}
\includegraphics[width=5.8 cm,height=5.cm,keepaspectratio]{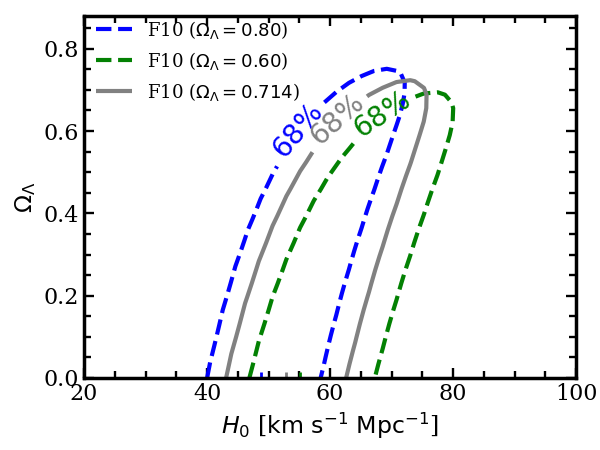}
\caption{Results from the sensitivity study of the Amati relation and cosmological parameters on the initial choice of $H_0$ and $\Omega_\Lambda$ to calculate $E_{\rm iso}$. The grey lines correspond to initial $\Omega_{\Lambda} = 0.714$  (default case).  The blue and green lines correspond to initial values of $\Omega_{\Lambda} = 0.8$ and 0.6, respectively.  The initial value of $H_0 = 69.6$~km~s$^{-1}$~Mpc$^{-1}$ (default case) is the same for all cases.}
\label{fig:hubble_Omega}
\end{center}
\end{figure*}

\subsection{Sensitivity on the initial choice of cosmological parameters}
Since the calculation of $E_\mathrm{iso}$ requires assuming standard values of the cosmological parameters (see equations \ref{Eiso_sbolo} and \ref{eq_dL}), the choice of these initial values may bias the Amati relation. To address this issue of circular logic, we reanalyzed the data using initial values different than the default values of $H_0 = 69.6$~km~s$^{-1}$~Mpc$^{-1}$ and $\Omega_{\Lambda} = 0.714$, and we tested the stability of our results. In this analysis, we have used the F10 sample of 25 {\it Fermi} GRBs as an example.

The results of our reanalysis are shown in Figs.~\ref{fig:hubble_H0} and \ref{fig:hubble_Omega}. First, in Fig.~\ref{fig:hubble_H0} we fix $\Omega_{\Lambda} = 0.714$ at its default value and vary $H_{0}$ to take values of 100~km~s$^{-1}$~Mpc$^{-1}$ (blue lines) and 50~km~s$^{-1}$~Mpc$^{-1}$ (green lines). Next, in Fig.~\ref{fig:hubble_Omega} we fix $H_0 = 69.6$~km~s$^{-1}$~Mpc$^{-1}$ at its default value and vary $\Omega_{\Lambda}$ to take values of 0.80 (blue lines) and 0.60 (green lines).  The default case with $H_0 = 69.6$~km~s$^{-1}$~Mpc$^{-1}$ and $\Omega_{\Lambda} = 0.714$ is shown as grey lines in both figures.

We find that the normalization parameter $k$ of the Amati relation is rather sensitive to the initial choice of $H_0$, while the slope parameter $m$ and the unknown systematic parameter $\sigma_{\rm ext}$ are virtually insensitive to this choice (see the left panel of Fig.~\ref{fig:hubble_H0}).  Therefore the Amati relation in the $E_\mathrm{i,p}$--$E_\mathrm{iso}$ plane just scales linearly with $H_0$ with the same slope (Fig.~\ref{fig:hubble_H0} middle panel).  The resulting $1\sigma$ contours in $\Omega_{\Lambda}$--$H_{0}$ plane also shifts to the higher values of $H_{0}$ with increasing initial values (Fig.~\ref{fig:hubble_H0} right panel).  On the other hand, the Amati relation parameters and unknown systematics parameter are only mildly sensitive to the initial choice of $\Omega_{\Lambda}$ (Fig.~\ref{fig:hubble_Omega} left panel).  As a result the shape of the Amati relation in the $E_\mathrm{i,p}$--$E_\mathrm{iso}$ plane is basically unchanged slope (Fig.~\ref{fig:hubble_Omega} middle panel) and the $1\sigma$ contours in $\Omega_{\Lambda}$--$H_{0}$ plane shifts mildly (Fig.~\ref{fig:hubble_Omega} right panel). The mild dependence on $\Omega_{\Lambda}$ results from the fact that the luminosity distance $d_L$ and hence $E_{\rm iso}$ depends more strongly on $H_0$ than $\Omega_{\Lambda}$.

\subsection{Joint fits with SNe U2.1 data}
Type-Ia supernovae can be standardized to use as cosmological standard candles and they provide constraints on the cosmological parameters at redshift $z < 2$~\citep{Perlmutter1999_517, Riess1998_116, Perlmutter_195}. In the meanwhile, type Ia SNe at maximum brightness appear to be better standard candles in the near infrared, requiring little or even without correction for the light-curve shape \citep{Krisciunas2004_602, Wood_Vasey2008_689}. However, it is necessary to use more distant objects to constrain these parameters. GRBs and SNe data together can potentially be used as a powerful tool for distance measurement and to probe the Hubble diagram at high redshifts. We have thus jointly analyzed GRBs in our samples together with the recent 580 SNe U2.1 sample from~\cite{Suzuki2012_746} that spans a redshift range from 0.015 to 1.414. For this combined dataset we simply estimate the best-fit parameters as the sum of both samples, i.e., $\chi^2_{\rm total} = \chi^2_{\rm GRBs} + \chi^2_{\rm SNe}$. 

\begin{figure}[htb]
\centering
\includegraphics[width=7.5 cm]{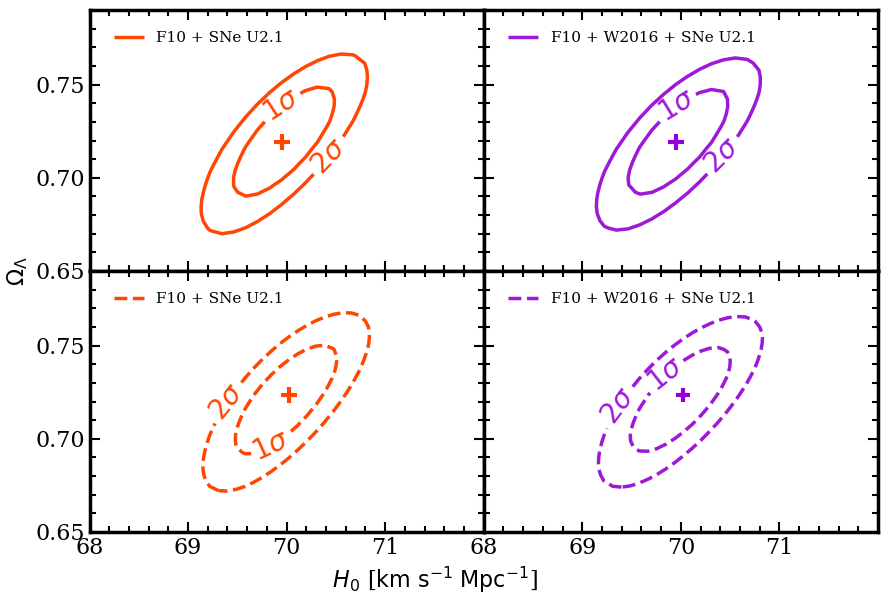}
\caption{The same as Fig.~\ref{fig:Cosmo_par1} but with the SNe U2.1 sample~\citep{Suzuki2012_746}. The top and bottom panels correspond to the GRBs samples with all redshifts and with redshifts $> 1.414$, respectively.}
\label{fig:Cosmo_par2}
\end{figure}

\begin{figure}[htb]
\begin{center}
\includegraphics[width=10.5 cm,height=5.cm,keepaspectratio]{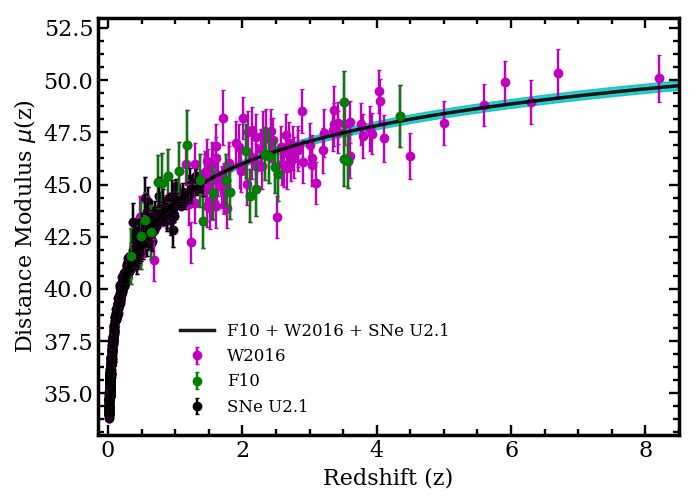}
\includegraphics[width=10.5 cm,height=5.cm,keepaspectratio]{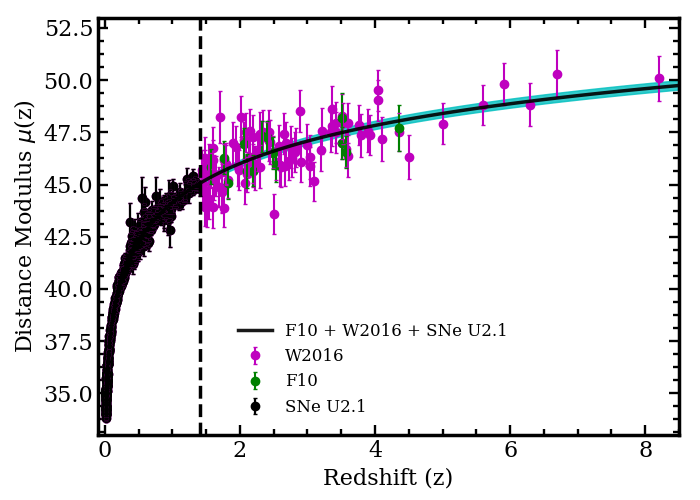}
\caption{Combined SNe and GRB Hubble diagram for the F10\,+\,W2016\,+\,SNe U2.1 sample. The top panel includes all GRBs and the bottom panel includes GRBs with $z>1.414$, respectively. The black lines are plotted using the estimated cosmological parameters obtained from the joint F10\,+\,W2016\,+\,SNe U2.1 data with 1$\sigma$ confidence regions (cyan), reported in Table~\ref{tab:cosmological_parameters}. The vertical broken line indicate $z = 1.414$.}
\label{fig:hubble_diagram}
\end{center}
\end{figure}

The results of this joint analysis are shown in Figs.~\ref{fig:Cosmo_par2} and \ref{fig:hubble_diagram}, and are also listed in Table~\ref{tab:cosmological_parameters}. Since the SNe U2.1 sample is much larger, the Hubble parameter $H_0$ and the density parameter $\Omega_\Lambda$ obtained from F10 + SNe U2.1 and F10 + W2016 + SNe U2.1 are well consistent with those obtained from the SNe U2.1 data alone, and are in agreement with the conclusion of~\cite{Wang2016_585, Demianski2017_693}. The $1\sigma$ and $2\sigma$ contours in the $H_0-\Omega_\Lambda$ plane, however, become much tighter (see Fig.~\ref{fig:Cosmo_par2}) in these cases thanks to the much larger weight of the SNe U2.1 sample. Figure~\ref{fig:hubble_diagram} shows the Hubble diagram constructed with the SNe U2.1 together with F10 and W2016 samples. The top panel includes all GRBs and the bottom panel includes GRBs with $z>1.414$, respectively.  The black solid line represents the distance moduli $\mu(z)$ obtained with the best-fit cosmological parameters obtained from the respective joint analyses.

\section{Discussion and Summary}\label{sec_6}
We have performed detailed time-integrated spectral analysis of GRBs detected with \textit{Fermi}-LAT and GBM in 9 years of operation and for which the redshift is known.  We found that the Band model \citep{Band1993_413} provides the best fit for 12 out of 26 GRBs in our sample while the SBPL model \citep{Ryde1999_39} provides the best fit for 4 GRBs (see Table~\ref{tab:modelfit} and Fig.~\ref{fig:models}).  Fitting the spectra of other GRBs require BB and/or PL and/or CPL model(s) in addition to the Band or SBPL model. The resulting $E_{\rm i,p}$ of the $\nu F_\nu$ spectra of these GRBs (see Table~\ref{tab:Sbolo}) varies from $60 \pm 2$~keV (GRB\,091127) to $19334 \pm 652$~keV (GRB\,160509A).  The corresponding isotropic-equivalent radiated energy from these bursts within $T_{90}$ ranges from $(1.3\pm 0.1)\times 10^{52}$~erg (GRB\,150514A) to $(435 \pm 6)\times 10^{52}$~erg (GRB\,160625B) in the 1--$10^4$~keV energy band $E_{\rm iso}^*$(F10).  Extending the energy band to 1--$10^5$~keV $E_{\rm iso}^{*}$(F100), we obtain different minimum and maximum values of $(1.4\pm 0.1)\times 10^{52}$~erg (GRB\,150514A) to $(605 \pm 25)\times 10^{52}$~erg (GRB\,080916C), respectively.  We used standard cosmological parameters, consistent with {\it Planck} results, to calculate these isotropic-equivalent energies. 

\textit{Fermi}-LAT samples the brightest and most luminous GRB population \citep{2013LATGRBCAT}.  This is clearly seen in the fluence and isotropic-equivalent energy distributions in Fig.~\ref{EisoSboloHistogram}.  Another recent GRB sample, W2016, used for investigating the Amati relation has lower average fluence and $E_{\rm iso}$ compared to our \textit{Fermi} sample.  The $E_{\rm i,p}$ in our sample is also generally higher than the W2016 sample.  Therefore, by selecting the \textit{Fermi}-LAT GRBs we are probing the $E_{\rm i,p}$--$E_{\rm iso}$ correlation for the brightest GRBs.  Besides, the analysis of these bright GRBs, often showing multiple spectral components, allow us to extend the computation of $E_{\rm iso}$ in the 1 keV -- 100 MeV energy range. To our knowledge this is the first study of the Amati relation with detailed spectral modeling using 9 years of \textit{Fermi} data.  We have also performed fits to the Amati relation using joint \textit{Fermi} + W2016 samples and have also explored its redshift dependence.

We find that the $E_{\rm i,p}$--$E_{\rm iso}$ correlation for the \textit{Fermi} LGRBs is rather strong and only the short GRB\,090510 appears as outlier (see Fig.~\ref{Cosmolo} and Table~\ref{tab:fitlkhood_2}). We did not use this burst in the rest of our analysis. The best-fit Amati relation for LGRBs can be expressed as 
$$
\frac{E_{\rm iso}}{10^{52}~{\rm erg}} = 
10^{1.67\pm 0.16} \left( \frac{E_{\rm i,p}}{950~{\rm keV}} \right)^{1.16\pm 0.37}\,,
$$
where $E_{\rm iso}$ is calculated in the 1--$10^4$~keV energy band. It must be stressed that the accuracy of the fits to the Amati relation is limited by our ignorance of its physical origin. Indeed, a full standardization of GRBs would require a better understanding of the non statistical scattering of the GRBs positions in the $E_{\rm i,p}$--$E_{\rm iso}$ plane. In our analysis, we thus used an ad-hoc uncertainty $\sigma_{\rm ext}$ (see equations \ref{DAgostini2005} and \ref{sigma_Eiso}) that represents these hidden physical parameters.   
The slope parameter by \cite{Wang2016_585} is $m = 1.48 \pm 0.09$. This value is consistent with our values of $m = 1.16 \pm 0.37$ (F10) and $m = 1.25 \pm 0.33$ (F100) within errors. The slope parameter $m = 1.69^{+0.07}_{-0.05}$ by \cite{Demianski2017_693} is slightly outside the 1$\sigma$ uncertainties.   
We reiterate that the parameter $k$ is not expected to be the same between F10 (or F100) and W2016 because the decorrelation energy is not the same between the samples (see Table \ref{tab:fitlkhood_2}). 

Our fit to the W2016 sample (94 LGRBs) and the joint \textit{Fermi} + W2016 sample (119 LGRBs) also resulted in parameter values compatible with other fits, although the errors on the parameters for these samples are smaller due to larger sample sizes. There are 14 LGRBs in the \textit{Fermi} sample with redshift $z>1.414$, which is the redshift up to which SNe type Ia have been observed. We could not obtain a good fit to the Amati relation using this high-redshift sub-sample, because of the small sub-sample size (see Fig.~\ref{fig:Cosmolo_1}). It is interesting to note however, that almost half of the \textit{Fermi}-LAT detected GRBs are at redshift $z<1.414$.  By combining the high-redshift GRBs from the \textit{Fermi} and W2016 samples (total 98 GRBs at $z>1.414$) we could obtain a good fit to the Amati relation.  The resulting fit parameters are similar to the parameters obtained from the complete \textit{Fermi} + W2016 sample, thus we see no evidence for an evolution of the best-fit Amati correlation parameters as a function of redshift.  

Our work confirms the Amati relation using a new set of GRBs with very well measured prompt emission spectra, and our analysis results are fully compatible with \cite{Wang2016_585}. Besides, when adding the F10 GRBs to the other sets of observed GRBs contained in the W2016 sample, we slightly improve the accuracy on the Amati relation parameter $m$ (from 0.20 to 0.16, see Table \ref{tab:fitlkhood_2}). However, our work also confirms that the extrinsic parameter is dominant in all analyses of this kind.

We have used a $\chi^2$ estimator to constrain the Hubble parameter $H_0$ and dark energy density $\Omega_\Lambda$ in a flat $\Lambda$CDM Universe by using the distance modulus $\mu = 5\log (d_{\rm L}/{\rm 1\,Mpc}) + 25$, where the luminosity distance $d_{\rm L}$ can be expressed in terms of $E_{\rm i,p}$ and $E_{\rm iso}$ through the Amati relation.  Thus GRBs are assumed standard cosmological candles, following the Amati relation fits.  This method has been used by many authors \citep[see, e.g., ][]{Amati2008_391, Wang2016_585, Demianski2017_693}, however, is somewhat circular in the sense that $E_{\rm iso}$ is calculated assuming particular values of $H_0$ and $\Omega_\Lambda$ in the first place.  Nevertheless we follow this method adopted by previous authors for illustration purpose and to explore the sensitivity of the Amati relation to cosmological parameters.  We find that the normalization of the Amati relation depends strongly on the initial choice of $H_0$ and is linearly proportional to it.  On the other hand, the initial choice of $\Omega_\Lambda$ has mild effect on both the normalization and slope of the Amati relation. In principle, the circularity problem could be solved by fitting the Amati relation and cosmological parameters simultaneously. Unfortunately, this simultaneous fitting method has not been very successful so far~\citep{ghirlanda2009advances, Wang2016_585}. A large sample of GRBs with less scatter in the $E_{\rm iso}$--$E_{\mathrm{i,p}}$ correlation will be useful in future for this study.

Because of a small sample size and of its limited range in redshift, the \textit{Fermi} GRBs cannot constrain the cosmological parameters (see Table \ref{tab:cosmological_parameters} and Fig.~\ref{fig:Cosmo_par1}). Our analysis of the W2016 sample gives best-fit values of $H_0$ and $\Omega_\Lambda$ similar to {\it Planck} results but no meaningful conclusions can be drawn because the errors are very large. The analysis of {\it Fermi}, W2016 and the joint {\it Fermi} + W2016 samples poorly constrain $H_0$ and $\Omega_\Lambda$ with large errors using both the whole samples and GRBs with $z>1.414$. Next, we have combined GRB samples ({\it Fermi} and W2016) with the SNe U2.1 sample \citep{Suzuki2012_746} and obtained the cosmological parameters (see Fig.~\ref{fig:Cosmo_par2}) as
$$
H_0 = 69.95^{+0.53}_{-0.47} \,\,;\,\, \Omega_\Lambda = 0.72\pm 0.03 \,.
$$
We have also fitted cosmological parameters using GRBs with $z>1.414$ and SNe U2.1 sample obtaining 
$$
H_0 = 70.03^{+0.46}_{-0.54} \,\,;\,\, \Omega_\Lambda = 0.72\pm 0.03 \,.
$$
Subsequently we have plotted the Hubble diagram up to redshift $z=8.2$ using these best-fit values, together with the GRB and SNe data (see Fig.~\ref{fig:hubble_diagram}).

GRBs have the potential to trace cosmological parameters up to high redshift, if they are standard cosmological candles.  The phenomenological Amati relation provides a window to explore this idea.  At present the quality of the correlation is poor to make big impact in cosmological study with GRBs but possibilities remain open with future big data sets.

\section*{Acknowledgments}
\noindent
The \textit{Fermi}-LAT Collaboration acknowledges support for LAT development, operation and data analysis from NASA and DOE (United States), CEA/Irfuand IN2P3/CNRS (France), ASI and INFN (Italy), MEXT, KEK, and JAXA (Japan), and the K.A. Wallenberg Foundation, the Swedish Research Council and the National Space Board (Sweden). Science analysis support in the operations phase from INAF (Italy) and CNES (France) is also gratefully acknowledged. The work presented in this paper was supported in part by a grant from the National Research Foundation (South Africa) and ORCID No. 0000-0002-3909-6711. F.~Dirirsa also acknowledges support from an Erasmus fellowship to visit Laboratoire Univers et Particules de Montpellier, Universit\'{e} Montpellier, CNRS/IN2P3, F-34095 Montpellier, France.

\newpage

\appendix

\section{Appendix A: Reanalysis of W2016 data} \label{W26_updated}

The data set analyzed by \citet{Wang2016_585} contains 151 entries, until GRB\,140213A, combining new data with old data from \citet{Amati2008_391} and \citet{Amati2009_508}.  Six {\it Fermi} GRBs which are common with our sample are GRB\,080916C, GRB\,090323, GRB\,090902B, GRB\,090926A,  GRB\,130427A and GRB\,130518A. A number of GRBs in \citet{Wang2016_585} are also counted twice.  Here we have reanalyzed 94 GRBs (W2016 sample) from this data set which are not in our {\it Fermi}-LAT/GBM GRB sample and are not counted twice.  The bolometric fluence $S_{\mathrm{bolo}}$ reported by \citet{Wang2016_585} has been converted to $E_{\mathrm{iso}}$ using the standard cosmological parameters and are reported in Table~\ref{my-label} together with $E_{\rm i,p}$. The $E_{\mathrm{iso}}$ has been calculated in the 1--$10^4$~keV range.

\begin{table}[h!]
\setlength{\tabcolsep}{0.3cm}
\parbox{.97\linewidth}{
\caption{The W2016 sample of 94 GRBs with $E_{\mathrm{iso}}$ updated from \citet{Amati2008_391,Amati2009_508, Wang2016_585}. \\ Instruments: SW = {\it Suzaku}-WAM,  KW = {\it Konus}-Wind~\citep{2017ApJ...850..161T}, SB = {\it Swift}-BAT, FG = {\it Fermi} GBM, HET = HETE-2, {\it GRO} = {\it CGRO}/BATSE and SAX = {\it Beppo}SAX.}
\centering
\scalebox{0.77}{
\begin{tabular}{lllllll}
\hline \hline
GRB &  Instruments  & z   & $E_{\mathrm{i,p}}$  & $S_{\mathrm{bolo}}$  & $E_{\mathrm{iso}}$ \\
  &    & &  (keV) &  (10$^{-5}$ erg/cm$^2$)  &  (10$^{52}$ erg)  & $\mu\pm\sigma_{\mu}$
\\ \hline
100413 & SW & 3.9    & 1783.60 $\pm$ 374.85 & 2.36 $\pm$ 0.77  & 72.80 $\pm$ 23.75 & 47.66 $\pm$ 1.07  \\ 
100621 & KW & 0.54   & 146.49 $\pm$ 23.9    & 5.75 $\pm$ 0.64  & 4.42 $\pm$ 0.49   & 42.54 $\pm$	1.01   \\
100704 & KW  & 3.6    & 809.60 $\pm$ 135.70  & 0.70 $\pm$ 0.07  & 19.00 $\pm$ 1.90  & 48.00 $\pm$ 1.01   \\
100814 & KW  & 1.44   & 312.32 $\pm$ 48.8    & 1.39 $\pm$ 0.23  & 7.59 $\pm$ 1.26   & 45.46 $\pm$ 1.01  \\
100906 & KW  & 1.73   & 387.23 $\pm$ 244.07  & 3.56 $\pm$ 0.55  & 27.36 $\pm$ 4.23  & 44.81 $\pm$ 1.24 \\
110205 & KW/SB/SW & 2.22   & 740.60 $\pm$ 322.0   & 3.32 $\pm$ 0.68  & 39.94 $\pm$ 8.18  & 45.82 $\pm$ 1.13 \\
110213 & KW  & 1.46   & 223.86 $\pm$ 70.11   & 1.55 $\pm$ 0.23  & 8.68 $\pm$ 1.29   & 44.97 $\pm$ 1.06  \\
110422 & KW  & 1.77   & 421.04 $\pm$ 13.85   & 9.32 $\pm$ 0.02  & 74.68 $\pm$ 0.16  & 43.88 $\pm$ 0.98   \\
110503 & KW  & 1.61   & 572.25 $\pm$ 50.95   & 2.76 $\pm$ 0.21  & 18.57 $\pm$ 1.41  & 45.49 $\pm$ 0.99   \\
110715 & KW  & 0.82   & 218.40 $\pm$ 20.93   & 2.73 $\pm$ 0.24  & 4.94 $\pm$ 0.43   & 44.00 $\pm$ 0.99   \\
110731 & KW  & 2.83   & 1164.32 $\pm$ 49.79  & 2.51 $\pm$ 0.01  & 45.85 $\pm$ 0.18  & 46.84 $\pm$ 0.98   \\
110818 & FG  & 3.36   & 1117.47 $\pm$ 241.11 & 1.05 $\pm$ 0.08  & 25.49 $\pm$ 1.94  & 47.88 $\pm$ 1.02   \\
111008 & KW  & 5.0     & 894.00 $\pm$ 240.0   & 1.06 $\pm$ 0.11  & 48.09 $\pm$ 4.99  & 47.95 $\pm$ 1.04  \\
111107 & FG  & 2.89   & 420.44 $\pm$ 124.58  & 0.18 $\pm$ 0.03  & 3.41 $\pm$ 0.57   & 48.53 $\pm$ 1.06   \\
111209 & KW  & 0.68   & 519.87 $\pm$ 88.88   & 69.47 $\pm$ 8.72 & 85.79 $\pm$ 10.77 & 41.40 $\pm$ 1.01   \\
120119 & KW  & 1.73   & 417.38 $\pm$ 54.56   & 4.62 $\pm$ 0.59  & 35.50 $\pm$ 4.53  & 44.62 $\pm$ 1.00   \\
120326 & FG  & 1.8    & 129.97 $\pm$ 10.27   & 0.44 $\pm$ 0.02  & 3.64 $\pm$ 0.17   & 45.85 $\pm$ 0.99 \\
120724 & SB  & 1.48   & 68.45 $\pm$ 18.60    & 0.15 $\pm$ 0.02  & 0.86 $\pm$ 0.11   & 46.14 $\pm$ 1.05 \\
120802 & SB  & 3.8    & 274.33 $\pm$ 93.04   & 0.43 $\pm$ 0.07  & 12.73 $\pm$ 2.07  & 47.32 $\pm$ 1.07   \\
120811C & SB & 2.67   & 157.49 $\pm$ 20.92   & 0.74 $\pm$ 0.07  & 12.25 $\pm$ 1.16 & 45.80 $\pm$ 1.00    \\
120909 & KW  & 3.93   & 1651.55 $\pm$ 123.25 & 2.69 $\pm$ 0.23  & 83.99 $\pm$ 7.18  & 47.44 $\pm$ 0.99   \\
120922 & SB  & 3.1    & 156.62 $\pm$ 0.04    & 1.59 $\pm$ 0.18  & 33.81 $\pm$ 3.83  & 45.08 $\pm$ 0.99   \\
121128 & KW  & 2.2    & 243.20 $\pm$ 12.8    & 0.87 $\pm$ 0.07  & 10.30 $\pm$ 0.83  & 45.98 $\pm$ 0.99   \\
130215 & FG  & 0.6    & 247.54 $\pm$ 100.61  & 4.84 $\pm$ 0.12  & 4.62 $\pm$ 0.11   & 43.38 $\pm$ 1.09   \\
130408 & KW  & 3.76   & 1003.94 $\pm$ 137.98 & 0.99 $\pm$ 0.17  & 28.81 $\pm$ 4.95  &  47.91 $\pm$ 1.01  \\
130420A & FG & 1.3    & 128.63 $\pm$ 6.89    & 1.73$\pm$ 0.06   & 7.77 $\pm$ 0.27  &  44.13 $\pm$ 0.99   \\
130505 & KW  & 2.27   & 2063.37 $\pm$ 101.37 & 4.56 $\pm$ 0.09  & 57.05 $\pm$ 1.13  &  46.68 $\pm$ 0.99  \\
130514 & KW/SB  & 3.6    & 496.80$\pm$ 151.8    & 1.88 $\pm$ 0.25  & 51.03 $\pm$ 6.79  &  46.36 $\pm$ 1.05  \\
130606 & KW  & 5.91   & 2031.54 $\pm$ 483.7  & 0.49 $\pm$ 0.09  & 28.56 $\pm$ 5.25  &  49.90 $\pm$ 1.04  \\
130610 & FG  & 2.09   & 911.83 $\pm$ 132.65  & 0.82 $\pm$ 0.05  & 8.87 $\pm$ 0.54   &  47.53 $\pm$ 1.00  \\
130612 & FG  & 2.01   & 186.07 $\pm$ 31.56   & 0.08 $\pm$ 0.01  & 0.81 $\pm$ 0.10    & 48.19 $\pm$ 1.01 \\
130701A & KW  & 1.16   & 191.80 $\pm$ 8.62    & 0.46 $\pm$ 0.04  & 1.66 $\pm$ 0.14   & 45.97 $\pm$ 0.99  \\
130831A & KW  & 0.48   & 81.35 $\pm$ 5.92     & 1.29 $\pm$ 0.07  & 0.78 $\pm$ 0.04   & 43.44 $\pm$ 0.99   \\
130907A & KW  & 1.24   & 881.77 $\pm$ 24.62   & 75.21 $\pm$ 4.76 & 308.4 $\pm$ 19.5 & 42.24 $\pm$ 0.99  \\
131030A & KW  & 1.29   & 405.86 $\pm$ 22.93   & 1.05 $\pm$ 0.10  & 4.65 $\pm$ 0.44   &  46.00 $\pm$ 0.99  \\
131105A & FG  & 1.69   & 547.68 $\pm$ 83.53   & 4.75 $\pm$ 0.16  & 34.96 $\pm$ 1.18  &  44.89 $\pm$ 1.00  \\
131117A & SB  & 4.04   & 221.85 $\pm$ 37.31   & 0.05 $\pm$ 0.01  & 1.63 $\pm$ 0.33   &  49.46 $\pm$ 1.03  \\
140206A & FG  & 2.73   & 447.60 $\pm$ 22.38   & 1.69 $\pm$ 0.03  & 29.05 $\pm$ 0.52  &  46.13 $\pm$ 0.98 \\
140213A & FG  & 1.21   & 176.61 $\pm$ 4.42    & 2.53 $\pm$ 0.04  & 9.89 $\pm$ 0.16   &  44.04 $\pm$ 0.98  \\ 
050318 & SB  & 1.44   & 115.00 $\pm$ 25.0    & 0.42 $\pm$ 0.03  & 2.29 $\pm$ 0.16    &  45.60 $\pm$ 1.02 \\
010222 & KW  & 1.48   & 766.00 $\pm$ 30.0    & 14.6 $\pm$ 1.50  & 83.92 $\pm$ 8.62   &  43.97 $\pm$ 0.99 \\
060418 & KW  & 1.489  & 572.00 $\pm$ 143.0   & 2.30 $\pm$ 0.50  & 13.37 $\pm$ 2.91   &  45.64 $\pm$ 1.05 \\
030328 & KW/HET & 1.52   & 328.00 $\pm$ 55.0    & 6.40 $\pm$ 0.60  & 38.68 $\pm$ 3.63   &  43.90 $\pm$ 1.01 \\
070125 & KW  & 1.547  & 934.00 $\pm$ 148.0   & 13.30 $\pm$ 1.30 & 83.07 $\pm$ 8.12   &  44.33 $\pm$ 1.01\\
090102 & KW  & 1.547  & 1149.00 $\pm$ 166.0  & 3.48 $\pm$ 0.63  & 21.74 $\pm$ 3.93   &  46.02 $\pm$ 1.02 \\
040912 & HET  & 1.563  & 44.00 $\pm$ 33.0     & 0.21 $\pm$ 0.06  & 1.34 $\pm$ 0.38    &  45.30 $\pm$ 1.35 \\
990123 & GRO/SAX/KW  & 1.6    & 1724.0 $\pm$ 466.0   & 35.80 $\pm$ 5.80 & 238.1 $\pm$ 38.58  &  43.98 $\pm$ 1.05 \\
071003 & KW  & 1.604  & 2077 $\pm$ 286  & 5.32 $\pm$ 0.590 & 35.55 $\pm$ 3.94  &  46.27 $\pm$ 1.01  \\
090418 & KW/SB  & 1.608  & 1567 $\pm$ 384   & 2.35 $\pm$ 0.59  & 15.78 $\pm$ 3.96   &  46.83 $\pm$ 1.06  \\
990510  & SAX	 & 1.619  & 423.0 $\pm$ 42.0     & 2.60 $\pm$ 0.40  & 17.68 $\pm$ 2.72   &  45.21 $\pm$ 1.00  \\
080605 & KW  	& 1.6398 & 650.0 $\pm$ 55.0     & 3.40 $\pm$ 0.28  & 23.67 $\pm$ 1.95   &  45.43 $\pm$ 0.99  \\
091020  & SB 	& 1.71   & 280.0 $\pm$ 190.0    & 0.11 $\pm$ 0.034 & 0.83 $\pm$ 0.26    &  48.20 $\pm$ 1.30 \\
080514B & KW  	& 1.8    & 627.0 $\pm$ 65.0    & 2.027 $\pm$ 0.48 & 16.75 $\pm$ 3.97   &  46.01 $\pm$ 1.02 \\
020127 & HET   & 1.9    & 290.0 $\pm$ 100.0    & 0.38 $\pm$ 0.01  & 3.46 $\pm$ 0.09    &  46.97 $\pm$ 1.06  \\
080319C & KW  	& 1.95   & 906.0 $\pm$ 272.0   & 1.50 $\pm$ 0.30  & 14.33 $\pm$ 2.87   &  46.82 $\pm$ 1.07  \\
081008 & SB  	& 1.9685 & 261.0 $\pm$ 52.0     & 0.96 $\pm$ 0.09  & 9.33 $\pm$ 0.87    &  45.87 $\pm$ 1.01 \\
030226  & HET   & 1.98   & 289.0 $\pm$ 66.0     & 1.30 $\pm$ 0.10  & 12.76 $\pm$ 0.98   &  45.66 $\pm$ 1.02 \\
000926  & KW  	& 2.07   & 310.0 $\pm$ 20.0     & 2.60 $\pm$ 0.60  & 27.64 $\pm$ 6.38   &  45.02 $\pm$ 1.02  \\
\hline
\end{tabular}}\\
\quad \quad Continued on next page\\
\label{my-label}
}
\end{table}

\begin{table}[h!]
\centering
\setlength{\tabcolsep}{0.3cm}
\parbox{.77\linewidth}{
\centering
\quad \quad Table \ref{my-label} – Continued from previous page\\
\scalebox{0.90}{
\begin{tabular}{lllllll}
\hline \hline
GRB & & z   & $E_{\mathrm{i,p}}$  & $S_{\mathrm{bolo}}$  & $E_{\mathrm{iso}}$ \\
  &  &  &  (keV) &  (10$^{-5}$ erg/cm$^2$)  &  (10$^{52}$ erg)  & $\mu\pm\sigma_{\mu}$
\\ \hline
011211  & SAX	& 2.14   & 186.0 $\pm$ 24.0     & 0.50 $\pm$ 0.06  & 5.64 $\pm$ 0.68    &  46.25 $\pm$ 1.00	\\
071020  & KW  	& 2.145 & 1013.0 $\pm$ 160.0  & 0.87 $\pm$ 0.40  & 9.85 $\pm$ 4.53     &  47.61 $\pm$ 1.12 	 \\ 
050922C & HET   & 2.198 & 415.0 $\pm$ 111.0   & 0.47 $\pm$ 0.16  & 5.56 $\pm$ 1.89     &  47.26 $\pm$ 1.09	\\
060124  & KW  	& 2.296 & 784.0 $\pm$ 285.0   & 3.40 $\pm$ 0.50  & 43.39 $\pm$ 6.38     &  45.89 $\pm$ 1.08	\\
021004  & HET   & 2.3   & 266.0 $\pm$ 117.0   & 0.27 $\pm$ 0.04  & 3.46 $\pm$ 0.51      &  47.38 $\pm$ 1.12	\\
051109A & KW  	& 2.346 & 539.0 $\pm$ 200.0   & 0.51 $\pm$ 0.05  & 6.76 $\pm$ 0.66 &   47.53 $\pm$ 1.08\\ 
060908 & SB  	& 2.43  & 514.0 $\pm$ 102.0   & 0.73 $\pm$ 0.07  & 10.28 $\pm$ 0.99 &  	47.11 $\pm$ 1.01\\
080413  & SW/SB & 2.433 & 584.0 $\pm$ 180.0   & 0.56 $\pm$ 0.14  & 7.90 $\pm$ 1.98  &  	47.55 $\pm$ 1.08 \\
090812   & SB 	& 2.452 & 2000.0 $\pm$ 700.0  & 3.08 $\pm$ 0.53  & 44.02 $\pm$ 7.58 &  	47.13 $\pm$ 1.08\\
081121  & KW  	& 2.512 & 47.23 $\pm$ 1.08    & 1.71 $\pm$ 0.33  & 25.51 $\pm$ 4.92 &  43.44 $\pm$ 1.01 	\\
081118  & SB/FG & 2.58  & 147.0 $\pm$ 14.0    & 0.27 $\pm$ 0.057 & 4.22 $\pm$ 0.89  &  46.79 $\pm$ 1.02 	\\
080721  & KW   & 2.591 & 1741.0 $\pm$ 227.0  & 7.86 $\pm$ 1.37  & 123.6 $\pm$ 21.6 &  45.99 $\pm$ 1.01 \\
050820   & KW   & 2.612 & 1325.0 $\pm$ 277.0  & 6.40 $\pm$ 0.50  & 102.06 $\pm$ 7.97 & 45.91 $\pm$ 1.02	\\
030429   & HET   & 2.65  & 128.0 $\pm$ 26.0    & 0.14 $\pm$ 0.02  & 2.29 $\pm$ 0.33   & 	47.36 $\pm$ 1.02\\
080603B  & KW   & 2.69  & 376.0 $\pm$ 100.0   & 0.64 $\pm$ 0.058 & 10.73 $\pm$ 0.97 &  46.97 $\pm$ 1.03\\
091029   & SB   & 2.752 & 230.0 $\pm$ 66.0    & 0.47 $\pm$ 0.044 & 8.19 $\pm$ 0.77   & 	46.75 $\pm$ 1.04 \\
081222  & FG   & 2.77  & 505.0 $\pm$ 34.0    & 1.67 $\pm$ 0.17  & 29.42 $\pm$ 3.00  & 	46.29 $\pm$ 0.99\\
050603   & KW   & 2.821 & 1333.0 $\pm$ 107.0  & 3.50 $\pm$ 0.20  & 63.59 $\pm$ 3.63  & 46.63 $\pm$ 0.99	\\
050401   & KW   & 2.9   & 467.0 $\pm$ 110.0   & 1.90 $\pm$ 0.40  & 36.16 $\pm$ 7.61  & 	46.10 $\pm$ 1.05\\
090715B  & SB   & 3.0   & 536.0 $\pm$ 172.0   & 1.09 $\pm$ 0.17  & 21.95 $\pm$ 3.42 & 46.89 $\pm$ 1.06 \\
080607  & KW   & 3.036 & 1691.0 $\pm$ 226.0  & 8.96 $\pm$ 0.48  & 184.07 $\pm$ 9.86 & 45.94 $\pm$ 1.00	\\
081028  & SB   & 3.038 & 234.0 $\pm$ 93.0    & 0.81 $\pm$ 0.095 & 16.66 $\pm$ 1.95  & 	46.26 $\pm$ 1.09\\
020124   & HET/KW  & 3.2   & 448.0 $\pm$ 148.0   & 1.20 $\pm$ 0.10  & 26.89 $\pm$ 2.24  & 	46.63 $\pm$ 1.06\\
060526   & SB 	& 3.21  & 105.0 $\pm$ 21.0    & 0.12 $\pm$ 0.06  & 2.70 $\pm$ 1.35   & 47.45 $\pm$ 1.15	\\
080810  & FG	 & 3.35  & 1470.0 $\pm$ 180.0  & 1.82 $\pm$ 0.20  & 43.96 $\pm$ 4.83  & 47.59 $\pm$ 1.00	\\
030323   & HET	& 3.37  & 270.0 $\pm$ 113.0   & 0.12 $\pm$ 0.04  & 2.93 $\pm$ 0.98   & 	48.59 $\pm$ 1.15\\
971214   & SAX	& 3.42  & 685.0 $\pm$ 133.0   & 0.87 $\pm$ 0.11  & 21.73 $\pm$ 2.75  & 	47.53 $\pm$ 1.02\\
060707   & SB 	& 3.425 & 279.0 $\pm$ 28.0    & 0.23 $\pm$ 0.04  & 5.76 $\pm$ 1.00   & 	47.93 $\pm$ 1.01\\
060115   & SB 	& 3.53  & 285.0 $\pm$ 34.0    & 0.25 $\pm$ 0.04  & 6.57 $\pm$ 1.05   &  47.89 $\pm$ 1.01\\
060206   & SB 	& 4.048 & 394.0 $\pm$ 46.0    & 0.14 $\pm$ 0.03  & 4.58 $\pm$ 0.98   & 	49.01 $\pm$ 1.02 \\
090516   & SB	& 4.109 & 971.0 $\pm$ 390.0   & 1.96 $\pm$ 0.38  & 65.66 $\pm$ 12.73 & 	47.21 $\pm$ 1.11\\
000131   & {\it GRO}/KW & 4.5   & 987.0 $\pm$ 416.0   & 4.70 $\pm$ 0.80  & 181.4 $\pm$ 30.9 & 46.36 $\pm$ 1.11\\
060927   & SB 	& 5.6   & 475.0 $\pm$ 47.0    & 0.27 $\pm$ 0.04  & 14.53 $\pm$ 2.15   & 48.81 $\pm$ 1.00\\
050904   & KW/SB & 6.29  & 3178 $\pm$ 1094.0 & 2.00 $\pm$ 0.20  & 127.7 $\pm$ 12.8 & 48.95 $\pm$ 1.07	\\
080913   & KW/SB & 6.695 & 710.0 $\pm$ 350.0   & 0.12 $\pm$ 0.035 & 8.39 $\pm$ 2.45    & 50.32 $\pm$ 1.18	\\
090423   & FG & 8.2   & 491.0 $\pm$ 200.0   & 0.12 $\pm$ 0.032 & 11.21 $\pm$ 2.99   & 50.09 $\pm$ 1.13\\ \hline
\end{tabular}}\\
\vspace{0.4cm}
}
\end{table}

\section{Appendix B: Spectral models}\label{spectral_models}

In this appendix we briefly describe the functional forms of different spectral models used in our analysis of the {\it Fermi} GRB data.  The best-fit model for individual GRBs is listed in Table~\ref{tab:modelfit}.

\subsection{Band function (Band)}
The Band function \citep{Band1993_413} is an empirical model that is widely used to fit the GRB prompt emission spectra.  It is the best-fit model for 12 GRBs in our {\it Fermi} sample and fits 6 other GRBs, together with one or more model(s). The Band function is composed of two power laws with indices $\alpha$ and $\beta$ joined by an exponential cutoff and amplitude $A$, in units of $\mathrm{{cm^{-2}~s^{-1}~keV^{-1}}}$, as given by
\begin{equation}
  {{N_{\mathrm{Band}}}(E)~\equiv~A}\begin{cases}
    \left({\frac{E}{100{~\mathrm{keV}}}}\right)^{{\alpha}} \exp\left[{-\frac{E(2+\alpha)}{E_{\mathrm{p}}}} \right]  & \mathrm{if}~ {E\leq E_{\mathrm{b}}}\\
    \left({\frac{E}{100{~\mathrm{keV}}}}\right)^{{\beta}}{\exp{(\beta-\alpha})}\left[{\frac{E_{\mathrm{p}}}{100{~\mathrm{keV}}}\frac{\alpha-\beta}{2+\alpha}} \right]^{\alpha-\beta}  & \mathrm{if}~ {E>E_{\mathrm{b}}} \,. 
  \end{cases}
\end{equation} 
Here $E_{\mathrm{p}}$ is the peak energy in keV of the $\nu F_\nu$ spectrum and ${E_{\mathrm{b}}= E_{\mathrm{p}}(\alpha-\beta)/(2+\alpha)}$ is the break energy.  In case of a very steep $\beta$, the high-energy part of the model is consistent with an exponential cutoff \citep{Kaneko2006_166}. The spectral peak energy in the cosmological rest-frame is $E_{\mathrm{i,p}} = E_{\mathrm{p}}(1+z)$.

\subsection{Power-law with an exponential cutoff (CPL)}
\label{CPL}
The power-law model with an exponential cutoff is given by
\begin{equation}
{{N_{\mathrm{CPL}}}(E) \equiv A\left(\frac{E}{100 ~\mathrm{keV}}\right)^{\gamma}\exp \left[-(2+\gamma)\frac{E}{E_{\mathrm{p}}}\right]}. 
\end{equation}
The model consists of 3 free parameters, namely the amplitude $A$, index $\gamma$ and peak energy $E_\mathrm{p}$. 
The intrinsic peak energy is given by $E_{\mathrm{i,p}} = E_{\mathrm{p}}(1+z)$.  None of the GRBs in our {\it Fermi} sample can be fitted with the CPL model alone, but the spectrum of GRB\,160509A is best fitted with Band+CPL+PL models.

\subsection{Smoothly-broken power law (SBPL)}
This is a broken power law with indices $\alpha$ and $\beta$ characterized by flexible curvature at the break energy \citep{Ryde1999_39}. Therefore this model can accommodate spectra with very sharp or smooth breaks. The functional form is given by 
\begin{equation}
{{N_{\mathrm{SBPL}}}(E) \equiv A\left(\frac{E}{100~\mathrm{keV}}\right)^{\frac{\alpha+\beta}{2}}10^{(a-a_{\mathrm{p}})}}, 
\end{equation}
where 
$${a = \frac{1}{2}\sigma(\beta-\alpha) \displaystyle\ln\left(\displaystyle\frac{e^{r}+e^{-r}}{2} \right)} \,;\,
{a_{\mathrm{p}} = \frac{1}{2}\sigma(\beta-\alpha) \ln\left(\frac{e^{r_{\mathrm{p}}}+e^{-r_{\mathrm{p}}}}{2}  \right)}\,;$$
$$r =  \displaystyle\frac{\log{(E/E_{\mathrm{0}})}}{\sigma}\,; \, r_{\mathrm{p}} = \frac{\log{(100~\mathrm{keV}/E_{\mathrm{0}})}}{\sigma}\,.$$
Here ${E_0}$ is the e-folding energy and ${\sigma}$ is the break scale in decades of energy fixed at 0.3.  The peak of the ${\nu F_{\mathrm{\nu}}}$ spectrum is at ${E_{\mathrm{p}} = E_{\mathrm{0}}10^{\left(1/2\sigma \ln\left[({\alpha+2})/({-\beta-2})\right]\right)}}$.  The SBPL alone is the best-fit model for four GRBs in our {\it Fermi} sample, while it is the best-fit together with another model in case of one other GRBs.

\subsection{Power law (PL)} 
A single PL with 2 free parameters, is given by  
\begin{equation}
N_{\rm PL}(E) \equiv A \left(\frac{E}{100~\mathrm{keV}}\right)^{\alpha_1} \,, 
\end{equation}
where $A$ is the amplitude and $\alpha_1$ is the photon index.  An additional PL component is required for 4 GRBs in our sample.

\subsection{Black-body (BB)}
An additional BB component is required for modeling the spectra of 5 GRBs in our {\it Fermi} sample. The functional form is given by  
\begin{equation}
N_{\rm BB}(E) = \displaystyle\frac{AE^2}{\exp(E/kT)-1}, 
\end{equation}
where $A$ is the amplitude and $kT$ is the thermal temperature.

\vspace{0.5cm}     

\bibliography{References_v26}

\begin{thebibliography}{}
\expandafter\ifx\csname natexlab\endcsname\relax\def\natexlab#1{#1}\fi

\bibitem[{{Abbott} {et~al.}(2017){Abbott}, {Abbott}, {Abbott}, {Acernese},
  {Ackley}, {Adams}, {Adams}, {Addesso}, {Adhikari}, {Adya}, \&
  et~al.}]{GW1708172017}
{Abbott}, B.~P., {Abbott}, R., {Abbott}, T.~D., {et~al.} 2017, \apjl, 848, L12

\bibitem[{Abdo {et~al.}(2009)Abdo, Ackermann, Arimoto, Asano, Atwood, Axelsson,
  Baldini, Ballet, Band, Barbiellini, {et~al.}}]{Abdo2009}
Abdo, A.~A., Ackermann, M., Arimoto, M., {et~al.} 2009, Science

\bibitem[{{Ackermann} {et~al.}(2010){Ackermann}, {Asano}, {Atwood}, {Axelsson},
  {Baldini}, {Ballet}, {Barbiellini}, {Baring}, {Bastieri}, {Bechtol},
  {Bellazzini}, {Berenji}, {Bhat}, {Bissaldi}, {Blandford}, {Bloom},
  {Bonamente}, {Borgland}, {Bouvier}, {Bregeon}, {Brez}, {Briggs}, {Brigida},
  {Bruel}, {Buson}, {Caliandro}, {Cameron}, {Caraveo}, {Carrigan},
  {Casandjian}, {Cecchi}, {{\c C}elik}, {Charles}, {Chiang}, {Ciprini},
  {Claus}, {Cohen-Tanugi}, {Connaughton}, {Conrad}, {Dermer}, {de Palma},
  {Dingus}, {Silva}, {Drell}, {Dubois}, {Dumora}, {Farnier}, {Favuzzi},
  {Fegan}, {Finke}, {Focke}, {Frailis}, {Fukazawa}, {Fusco}, {Gargano},
  {Gasparrini}, {Gehrels}, {Germani}, {Giglietto}, {Giordano}, {Glanzman},
  {Godfrey}, {Granot}, {Grenier}, {Grondin}, {Grove}, {Guiriec}, {Hadasch},
  {Harding}, {Hays}, {Horan}, {Hughes}, {J{\'o}hannesson}, {Johnson}, {Kamae},
  {Katagiri}, {Kataoka}, {Kawai}, {Kippen}, {Kn{\"o}dlseder}, {Kocevski},
  {Kouveliotou}, {Kuss}, {Lande}, {Latronico}, {Lemoine-Goumard}, {Llena
  Garde}, {Longo}, {Loparco}, {Lott}, {Lovellette}, {Lubrano}, {Makeev},
  {Mazziotta}, {McEnery}, {McGlynn}, {Meegan}, {M{\'e}sz{\'a}ros}, {Michelson},
  {Mitthumsiri}, {Mizuno}, {Moiseev}, {Monte}, {Monzani}, {Moretti},
  {Morselli}, {Moskalenko}, {Murgia}, {Nakajima}, {Nakamori}, {Nolan},
  {Norris}, {Nuss}, {Ohno}, {Ohsugi}, {Omodei}, {Orlando}, {Ormes}, {Ozaki},
  {Paciesas}, {Paneque}, {Panetta}, {Parent}, {Pelassa}, {Pepe},
  {Pesce-Rollins}, {Piron}, {Preece}, {Rain{\`o}}, {Rando}, {Razzano},
  {Razzaque}, {Reimer}, {Ritz}, {Rodriguez}, {Roth}, {Ryde}, {Sadrozinski},
  {Sander}, {Scargle}, {Schalk}, {Sgr{\`o}}, {Siskind}, {Smith}, {Spandre},
  {Spinelli}, {Stamatikos}, {Stecker}, {Strickman}, {Suson}, {Tajima},
  {Takahashi}, {Takahashi}, {Tanaka}, {Thayer}, {Thayer}, {Thompson},
  {Tibaldo}, {Toma}, {Torres}, {Tosti}, {Tramacere}, {Uchiyama}, {Uehara},
  {Usher}, {van der Horst}, {Vasileiou}, {Vilchez}, {Vitale}, {von Kienlin},
  {Waite}, {Wang}, {Wilson-Hodge}, {Winer}, {Wu}, {Yamazaki}, {Yang}, {Ylinen},
  \& {Ziegler}}]{Ackermann2010_716}
{Ackermann}, M., {Asano}, K., {Atwood}, W.~B., {et~al.} 2010, The Astrophysical
  Journal, 716, 1178

\bibitem[{{Ackermann} {et~al.}(2013){Ackermann}, {Ajello}, {Asano}, {Axelsson},
  {Baldini}, {Ballet}, {Barbiellini}, {Bastieri}, {Bechtol}, {Bellazzini},
  {Bhat}, {Bissaldi}, {Bloom}, {Bonamente}, {Bonnell}, {Bouvier}, {Brandt},
  {Bregeon}, {Brigida}, {Bruel}, {Buehler}, {Burgess}, {Buson}, {Byrne},
  {Caliandro}, {Cameron}, {Caraveo}, {Cecchi}, {Charles}, {Chaves},
  {Chekhtman}, {Chiang}, {Chiaro}, {Ciprini}, {Claus}, {Cohen-Tanugi},
  {Connaughton}, {Conrad}, {Cutini}, {D'Ammando}, {de Angelis}, {de Palma},
  {Dermer}, {Desiante}, {Digel}, {Dingus}, {Di Venere}, {Drell},
  {Drlica-Wagner}, {Dubois}, {Favuzzi}, {Ferrara}, {Fitzpatrick}, {Foley},
  {Franckowiak}, {Fukazawa}, {Fusco}, {Gargano}, {Gasparrini}, {Gehrels},
  {Germani}, {Giglietto}, {Giommi}, {Giordano}, {Giroletti}, {Glanzman},
  {Godfrey}, {Goldstein}, {Granot}, {Grenier}, {Grove}, {Gruber}, {Guiriec},
  {Hadasch}, {Hanabata}, {Hayashida}, {Horan}, {Hou}, {Hughes}, {Inoue},
  {Jackson}, {Jogler}, {J{\'o}hannesson}, {Johnson}, {Johnson}, {Kamae},
  {Kataoka}, {Kawano}, {Kippen}, {Kn{\"o}dlseder}, {Kocevski}, {Kouveliotou},
  {Kuss}, {Lande}, {Larsson}, {Latronico}, {Lee}, {Longo}, {Loparco},
  {Lovellette}, {Lubrano}, {Massaro}, {Mayer}, {Mazziotta}, {McBreen},
  {McEnery}, {McGlynn}, {Michelson}, {Mizuno}, {Moiseev}, {Monte}, {Monzani},
  {Moretti}, {Morselli}, {Murgia}, {Nemmen}, {Nuss}, {Nymark}, {Ohno},
  {Ohsugi}, {Omodei}, {Orienti}, {Orlando}, {Paciesas}, {Paneque}, {Panetta},
  {Pelassa}, {Perkins}, {Pesce-Rollins}, {Piron}, {Pivato}, {Porter}, {Preece},
  {Racusin}, {Rain{\`o}}, {Rando}, {Rau}, {Razzano}, {Razzaque}, {Reimer},
  {Reimer}, {Reposeur}, {Ritz}, {Romoli}, {Roth}, {Ryde}, {Saz Parkinson},
  {Schalk}, {Sgr{\`o}}, {Siskind}, {Sonbas}, {Spandre}, {Spinelli}, {Suson},
  {Tajima}, {Takahashi}, {Takeuchi}, {Tanaka}, {Thayer}, {Thayer}, {Thompson},
  {Tibaldo}, {Tierney}, {Tinivella}, {Torres}, {Tosti}, {Troja}, {Tronconi},
  {Usher}, {Vandenbroucke}, {van der Horst}, {Vasileiou}, {Vianello}, {Vitale},
  {von Kienlin}, {Winer}, {Wood}, {Wood}, {Xiong}, \& {Yang}}]{2013LATGRBCAT}
{Ackermann}, M., {Ajello}, M., {Asano}, K., {et~al.} 2013, \apjs, 209, 11

\bibitem[{{Amati}(2006{\natexlab{a}})}]{Amati2006_121}
{Amati}, L. 2006{\natexlab{a}}, Nuovo Cimento B Serie, 121, 1081

\bibitem[{{Amati}(2006{\natexlab{b}})}]{Amati2006_372}
---. 2006{\natexlab{b}}, Monthly Notices of the Royal Astronomical Society,
  372, 233

\bibitem[{Amati {et~al.}(2009)Amati, Frontera, \& Guidorzi}]{Amati2009_508}
Amati, L., Frontera, F., \& Guidorzi, C. 2009, Astronomy \& Astrophysics, 508,
  173

\bibitem[{{Amati} {et~al.}(2008){Amati}, {Guidorzi}, {Frontera}, {Della Valle},
  {Finelli}, {Landi}, \& {Montanari}}]{Amati2008_391}
{Amati}, L., {Guidorzi}, C., {Frontera}, F., {et~al.} 2008, Monthly Notices of
  the Royal Astronomical Society, 391, 577

\bibitem[{{Amati} {et~al.}(2002){Amati}, {Frontera}, {Tavani}, {in't Zand},
  {Antonelli}, {Costa}, {Feroci}, {Guidorzi}, {Heise}, {Masetti}, {Montanari},
  {Nicastro}, {Palazzi}, {Pian}, {Piro}, \& {Soffitta}}]{Amati2002_81A}
{Amati}, L., {Frontera}, F., {Tavani}, M., {et~al.} 2002, Astronomy \&
  Astrophysics, 390, 81

\bibitem[{Atwood {et~al.}(2009)}]{Atwood2009_697}
Atwood, W., {et~al.} 2009, The Astrophysical Journal, 697, 1071

\bibitem[{{Axelsson} {et~al.}(2012){Axelsson}, {Baldini}, {Barbiellini},
  {Baring}, {Bellazzini}, {Bregeon}, {Brigida}, {Bruel}, {Buehler},
  {Caliandro}, {Cameron}, {Caraveo}, {Cecchi}, {Chaves}, {Chekhtman}, {Chiang},
  {Claus}, {Conrad}, {Cutini}, {D'Ammando}, {de Palma}, {Dermer}, {Silva},
  {Drell}, {Favuzzi}, {Fegan}, {Ferrara}, {Focke}, {Fukazawa}, {Fusco},
  {Gargano}, {Gasparrini}, {Gehrels}, {Germani}, {Giglietto}, {Giroletti},
  {Godfrey}, {Guiriec}, {Hadasch}, {Hanabata}, {Hayashida}, {Hou}, {Iyyani},
  {Jackson}, {Kocevski}, {Kuss}, {Larsson}, {Larsson}, {Longo}, {Loparco},
  {Lundman}, {Mazziotta}, {McEnery}, {Mizuno}, {Monzani}, {Moretti},
  {Morselli}, {Murgia}, {Nuss}, {Nymark}, {Ohno}, {Omodei}, {Pesce-Rollins},
  {Piron}, {Pivato}, {Racusin}, {Rain{\`o}}, {Razzano}, {Razzaque}, {Reimer},
  {Roth}, {Ryde}, {Sanchez}, {Sgr{\`o}}, {Siskind}, {Spandre}, {Spinelli},
  {Stamatikos}, {Tibaldo}, {Tinivella}, {Usher}, {Vandenbroucke}, {Vasileiou},
  {Vianello}, {Vitale}, {Waite}, {Winer}, {Wood}, {Burgess}, {Bhat},
  {Bissaldi}, {Briggs}, {Connaughton}, {Fishman}, {Fitzpatrick}, {Foley},
  {Gruber}, {Kippen}, {Kouveliotou}, {Jenke}, {McBreen}, {McGlynn}, {Meegan},
  {Paciesas}, {Pelassa}, {Preece}, {Tierney}, {von Kienlin}, {Wilson-Hodge},
  {Xiong}, \& {Pe'er}}]{Axelsson2012_757}
{Axelsson}, M., {Baldini}, L., {Barbiellini}, G., {et~al.} 2012, \apjl, 757,
  L31

\bibitem[{{Band} {et~al.}(1993){Band}, {Matteson}, {Ford}, {Schaefer},
  {Palmer}, {Teegarden}, {Cline}, {Briggs}, {Paciesas}, {Pendleton}, {Fishman},
  {Kouveliotou}, {Meegan}, {Wilson}, \& {Lestrade}}]{Band1993_413}
{Band}, D., {Matteson}, J., {Ford}, L., {et~al.} 1993, The Astrophysical
  Journal, 413, 281

\bibitem[{{Bennett} {et~al.}(2014){Bennett}, {Larson}, {Weiland}, \&
  {Hinshaw}}]{Bennett2014_794}
{Bennett}, C.~L., {Larson}, D., {Weiland}, J.~L., \& {Hinshaw}, G. 2014, \apj,
  794, 135

\bibitem[{{Berger}(2011)}]{110721A_3.512_Berger2011GCN_12193}
{Berger}, E. 2011, GRB Coordinates Network, Circular Service, No.~12193, \#1
  (2011), 12193

\bibitem[{{Butler} {et~al.}(2010){Butler}, {Bloom}, \&
  {Poznanski}}]{Butler2010}
{Butler}, N.~R., {Bloom}, J.~S., \& {Poznanski}, D. 2010, \apj, 711, 495

\bibitem[{{Butler} {et~al.}(2009){Butler}, {Kocevski}, \& {Bloom}}]{Butler2009}
{Butler}, N.~R., {Kocevski}, D., \& {Bloom}, J.~S. 2009, \apj, 694, 76

\bibitem[{{Butler} {et~al.}(2007){Butler}, {Kocevski}, {Bloom}, \&
  {Curtis}}]{Butler2007}
{Butler}, N.~R., {Kocevski}, D., {Bloom}, J.~S., \& {Curtis}, J.~L. 2007, \apj,
  671, 656

\bibitem[{{Cenko} {et~al.}(2009){Cenko}, {Bloom}, {Morgan}, \&
  {Perley}}]{090328_0.736_Cenko2009GCN_9053}
{Cenko}, S.~B., {Bloom}, J.~S., {Morgan}, A.~N., \& {Perley}, D.~A. 2009, GRB
  Coordinates Network, 9053

\bibitem[{{Chornock} {et~al.}(2009{\natexlab{a}}){Chornock}, {Perley}, {Cenko},
  \& {Bloom}}]{090323_3.57_Chornock2009GCN_9028}
{Chornock}, R., {Perley}, D.~A., {Cenko}, S.~B., \& {Bloom}, J.~S.
  2009{\natexlab{a}}, GRB Coordinates Network, 9028

\bibitem[{{Chornock} {et~al.}(2009{\natexlab{b}}){Chornock}, {Perley}, {Cenko},
  \& {Bloom}}]{090424_0.544_Chornock2009GCN_9243}
---. 2009{\natexlab{b}}, GRB Coordinates Network, 9243

\bibitem[{{Collazzi} {et~al.}(2012){Collazzi}, {Schaefer}, {Goldstein}, \&
  {Preece}}]{Collazzi2012}
{Collazzi}, A.~C., {Schaefer}, B.~E., {Goldstein}, A., \& {Preece}, R.~D. 2012,
  \apj, 747, 39

\bibitem[{{Cucchiara}(2010)}]{100414A_1.368_Cucchiara2010GCN_10608}
{Cucchiara}, A. 2010, GRB Coordinates Network, Circular Service, No.~10608, \#1
  (2010), 10608

\bibitem[{{Cucchiara}(2014)}]{131231A_0.6439_Cucchiara2014GCN_15652}
---. 2014, GRB Coordinates Network, Circular Service, No.~15652, \#1 (2014),
  15652

\bibitem[{{Cucchiara} {et~al.}(2009{\natexlab{a}}){Cucchiara}, {Fox}, {Levan},
  \& {Tanvir}}]{091127_0.49_Cucchiara2009GCN_10202}
{Cucchiara}, A., {Fox}, D., {Levan}, A., \& {Tanvir}, N. 2009{\natexlab{a}},
  GRB Coordinates Network, Circular Service, No.~10202, \#1 (2009), 10202

\bibitem[{{Cucchiara} {et~al.}(2009{\natexlab{b}}){Cucchiara}, {Fox}, {Cenko},
  {Tanvir}, \& {Berger}}]{091003_0.8969_Cucchiara2009GCN_10031}
{Cucchiara}, A., {Fox}, D.~B., {Cenko}, S.~B., {Tanvir}, N., \& {Berger}, E.
  2009{\natexlab{b}}, GRB Coordinates Network, Circular Service, No.~1031, \#1
  (2009), 10031

\bibitem[{{Cucchiara} {et~al.}(2009{\natexlab{c}}){Cucchiara}, {Fox}, {Tanvir},
  \& {Berger}}]{090902B_1.822_Cucchiara2009GCN_9873}
{Cucchiara}, A., {Fox}, D.~B., {Tanvir}, N., \& {Berger}, E.
  2009{\natexlab{c}}, GRB Coordinates Network, 9873

\bibitem[{{Cucchiara} {et~al.}(2011){Cucchiara}, {Levan}, {Fox}, {Tanvir},
  {Ukwatta}, {Berger}, {Kr{\"u}hler}, {K{\"u}pc{\"u} Yolda{\c s}}, {Wu},
  {Toma}, {Greiner}, {Olivares}, {Rowlinson}, {Amati}, {Sakamoto}, {Roth},
  {Stephens}, {Fritz}, {Fynbo}, {Hjorth}, {Malesani}, {Jakobsson}, {Wiersema},
  {O'Brien}, {Soderberg}, {Foley}, {Fruchter}, {Rhoads}, {Rutledge}, {Schmidt},
  {Dopita}, {Podsiadlowski}, {Willingale}, {Wolf}, {Kulkarni}, \&
  {D'Avanzo}}]{Cucchiara2011_736}
{Cucchiara}, A., {Levan}, A.~J., {Fox}, D.~B., {et~al.} 2011, The Astrophysical
  Journal, 736, 7

\bibitem[{{D'Agostini}(2005)}]{DAgostini2005_11182D}
{D'Agostini}, G. 2005, ArXiv Physics e-prints, physics/0511182

\bibitem[{{de Ugarte Postigo} {et~al.}(2017){de Ugarte Postigo}, {Kann},
  {Thoene}, {Izzo}, {Tanvir}, {Lombardi}, \&
  {Marante}}]{170405A_3.51_Postigo2017GCN_20990}
{de Ugarte Postigo}, A., {Kann}, D.~A., {Thoene}, C.~C., {et~al.} 2017, GRB
  Coordinates Network, Circular Service, No.~20990, \#1 (2017), 20990

\bibitem[{{de Ugarte Postigo} {et~al.}(2013{\natexlab{a}}){de Ugarte Postigo},
  {Thoene}, {Gorosabel}, {Sanchez-Ramirez}, {Fynbo}, {Tanvir},
  {Cabrera-Lavers}, \& {Garcia}}]{131108A_2.40_Postigo2013GCN_15470}
{de Ugarte Postigo}, A., {Thoene}, C.~C., {Gorosabel}, J., {et~al.}
  2013{\natexlab{a}}, GRB Coordinates Network, Circular Service, No.~15470, \#1
  (2013), 15470

\bibitem[{{de Ugarte Postigo} {et~al.}(2015{\natexlab{a}}){de Ugarte Postigo},
  {Xu}, {Malesani}, \& {Tanvir}}]{150514A_0.807_Postigo2015GCN_17822}
{de Ugarte Postigo}, A., {Xu}, D., {Malesani}, D., \& {Tanvir}, N.~R.
  2015{\natexlab{a}}, GRB Coordinates Network, Circular Service, No.~17822, \#1
  (2015), 17822

\bibitem[{{de Ugarte Postigo} {et~al.}(2013{\natexlab{b}}){de Ugarte Postigo},
  {Campana}, {Th{\"o}ne}, {D'Avanzo}, {S{\'a}nchez-Ram{\'{\i}}rez}, {Melandri},
  {Gorosabel}, {Ghirlanda}, {Veres}, {Mart{\'{\i}}n}, {Petitpas}, {Covino},
  {Fynbo}, \& {Levan}}]{120624B_2.2_Postigo2013_557}
{de Ugarte Postigo}, A., {Campana}, S., {Th{\"o}ne}, C.~C., {et~al.}
  2013{\natexlab{b}}, Astronomy \& Astrophysics, 557, L18

\bibitem[{{de Ugarte Postigo} {et~al.}(2015{\natexlab{b}}){de Ugarte Postigo},
  {Fynbo}, {Thoene}, {Tanvir}, {Sanchez-Ramirez}, {Gorosabel}, {Pessev},
  {Alvarez-Iglesias}, \& {Rivero}}]{150314A_1.758_Postigo2015GCN_17583}
{de Ugarte Postigo}, A., {Fynbo}, J.~P.~U., {Thoene}, C., {et~al.}
  2015{\natexlab{b}}, GRB Coordinates Network, Circular Service, No.~17583, \#1
  (2015), 17583

\bibitem[{{Demianski} \& {Piedipalumbo}(2011)}]{Demianski2011_415}
{Demianski}, M., \& {Piedipalumbo}, E. 2011, \mnras, 415, 3580

\bibitem[{Demianski {et~al.}(2017)Demianski, Piedipalumbo, Sawant, \&
  Amati}]{Demianski2017_693}
Demianski, M., Piedipalumbo, E., Sawant, D., \& Amati, L. 2017, Astronomy \&
  Astrophysics, 598, A112

\bibitem[{{Dirirsa} \& {Razzaque}(2017)}]{Dirirsa2017}
{Dirirsa}, F.~F., \& {Razzaque}, S. 2017, in 5th Annual Conference on High
  Energy Astrophysics in Southern Africa, 2

\bibitem[{{Eichler} {et~al.}(1989){Eichler}, {Livio}, {Piran}, \&
  {Schramm}}]{Eichler1989}
{Eichler}, D., {Livio}, M., {Piran}, T., \& {Schramm}, D.~N. 1989, \nat, 340,
  126

\bibitem[{{Fishman} \& {Meegan}(1995)}]{Fishman1995}
{Fishman}, G.~J., \& {Meegan}, C.~A. 1995, \araa, 33, 415

\bibitem[{{Gehrels} \& {Razzaque}(2013)}]{Gehrels2013}
{Gehrels}, N., \& {Razzaque}, S. 2013, Frontiers of Physics, 8, 661

\bibitem[{Ghirlanda(2009)}]{ghirlanda2009advances}
Ghirlanda, G. 2009in , AIP, 579--586

\bibitem[{{Ghirlanda} {et~al.}(2004){Ghirlanda}, {Ghisellini}, \&
  {Lazzati}}]{Ghirlanda2004_616}
{Ghirlanda}, G., {Ghisellini}, G., \& {Lazzati}, D. 2004, \apj, 616, 331

\bibitem[{Ghirlanda {et~al.}(2009)Ghirlanda, Nava, Ghisellini, Celotti, \&
  Firmani}]{Ghirlanda2009_496}
Ghirlanda, G., Nava, L., Ghisellini, G., Celotti, A., \& Firmani, C. 2009,
  Astronomy \& Astrophysics, 496, 585

\bibitem[{{Ghirlanda} {et~al.}(2008){Ghirlanda}, {Nava}, {Ghisellini},
  {Firmani}, \& {Cabrera}}]{Ghirlanda2008_387}
{Ghirlanda}, G., {Nava}, L., {Ghisellini}, G., {Firmani}, C., \& {Cabrera},
  J.~I. 2008, Monthly Notices of the Royal Astronomical Society, 387, 319

\bibitem[{{Goldstein} {et~al.}(2017){Goldstein}, {Veres}, {Burns}, {Briggs},
  {Hamburg}, {Kocevski}, {Wilson-Hodge}, {Preece}, {Poolakkil}, {Roberts},
  {Hui}, {Connaughton}, {Racusin}, {von Kienlin}, {Dal Canton}, {Christensen},
  {Littenberg}, {Siellez}, {Blackburn}, {Broida}, {Bissaldi}, {Cleveland},
  {Gibby}, {Giles}, {Kippen}, {McBreen}, {McEnery}, {Meegan}, {Paciesas}, \&
  {Stanbro}}]{GRB170817A}
{Goldstein}, A., {Veres}, P., {Burns}, E., {et~al.} 2017, \apjl, 848, L14

\bibitem[{{Gonz{\'a}lez} {et~al.}(2003){Gonz{\'a}lez}, {Dingus}, {Kaneko},
  {Preece}, {Dermer}, \& {Briggs}}]{Gonzalex2003_424}
{Gonz{\'a}lez}, M.~M., {Dingus}, B.~L., {Kaneko}, Y., {et~al.} 2003, \nat, 424,
  749

\bibitem[{{Greiner} {et~al.}(2009){Greiner}, {Clemens}, {Kr{\"u}hler}, {von
  Kienlin}, {Rau}, {Sari}, {Fox}, {Kawai}, {Afonso}, {Ajello}, {Berger},
  {Cenko}, {Cucchiara}, {Filgas}, {Klose}, {K{\"u}pc{\"u} Yolda{\c s}},
  {Lichti}, {L{\"o}w}, {McBreen}, {Nagayama}, {Rossi}, {Sato}, {Szokoly},
  {Yolda{\c s}}, \& {Zhang}}]{Greiner2009_498}
{Greiner}, J., {Clemens}, C., {Kr{\"u}hler}, T., {et~al.} 2009, Astronomy \&
  Astrophysics, 498, 89

\bibitem[{{Guiriec} {et~al.}(2017){Guiriec}, {Gehrels}, {McEnery},
  {Kouveliotou}, \& {Hartmann}}]{Guiriec2017_846}
{Guiriec}, S., {Gehrels}, N., {McEnery}, J., {Kouveliotou}, C., \& {Hartmann},
  D.~H. 2017, \apj, 846, 138

\bibitem[{{Guiriec} {et~al.}(2016){Guiriec}, {Gonzalez}, {Sacahui},
  {Kouveliotou}, {Gehrels}, \& {McEnery}}]{Guiriec2016_819}
{Guiriec}, S., {Gonzalez}, M.~M., {Sacahui}, J.~R., {et~al.} 2016, \apj, 819,
  79

\bibitem[{{Guiriec} {et~al.}(2010){Guiriec}, {Briggs}, {Connaugthon}, {Kara},
  {Daigne}, {Kouveliotou}, {van der Horst}, {Paciesas}, {Meegan}, {Bhat},
  {Foley}, {Bissaldi}, {Burgess}, {Chaplin}, {Diehl}, {Fishman}, {Gibby},
  {Giles}, {Goldstein}, {Greiner}, {Gruber}, {von Kienlin}, {Kippen},
  {McBreen}, {Preece}, {Rau}, {Tierney}, \& {Wilson-Hodge}}]{Guiriec2010_725}
{Guiriec}, S., {Briggs}, M.~S., {Connaugthon}, V., {et~al.} 2010, \apj, 725,
  225

\bibitem[{Guiriec {et~al.}(2011)Guiriec, Connaughton, Briggs, Burgess, Ryde,
  Daigne, Mészáros, Goldstein, McEnery, Omodei, Bhat, Bissaldi,
  Camero-Arranz, Chaplin, Diehl, Fishman, Foley, Gibby, Giles, Greiner, Gruber,
  von Kienlin, Kippen, Kouveliotou, McBreen, Meegan, Paciesas, Preece, Rau,
  Tierney, van~der Horst, \& Wilson-Hodge}]{Guiriec2011_727}
Guiriec, S., Connaughton, V., Briggs, M.~S., {et~al.} 2011, The Astrophysical
  Journal Letters, 727, L33

\bibitem[{{Guiriec} {et~al.}(2013){Guiriec}, {Daigne}, {Hasco{\"e}t},
  {Vianello}, {Ryde}, {Mochkovitch}, {Kouveliotou}, {Xiong}, {Bhat}, {Foley},
  {Gruber}, {Burgess}, {McGlynn}, {McEnery}, \& {Gehrels}}]{Guiriec2013_770}
{Guiriec}, S., {Daigne}, F., {Hasco{\"e}t}, R., {et~al.} 2013, The
  Astrophysical Journal, 770, 32

\bibitem[{{Guiriec} {et~al.}(2015){Guiriec}, {Kouveliotou}, {Daigne}, {Zhang},
  {Hasco{\"e}t}, {Nemmen}, {Thompson}, {Bhat}, {Gehrels}, {Gonzalez}, {Kaneko},
  {McEnery}, {Mochkovitch}, {Racusin}, {Ryde}, {Sacahui}, \&
  {{\"U}nsal}}]{Guiriec2015_807}
{Guiriec}, S., {Kouveliotou}, C., {Daigne}, F., {et~al.} 2015, The
  Astrophysical Journal, 807, 148

\bibitem[{{Heussaff} {et~al.}(2013){Heussaff}, {Atteia}, \&
  {Zolnierowski}}]{Heussaff2013_557A}
{Heussaff}, V., {Atteia}, J.-L., \& {Zolnierowski}, Y. 2013, Astronomy \&
  Astrophysics, 557, A100

\bibitem[{Kaneko {et~al.}(2006)Kaneko, Preece, Briggs, Paciesas, Meegan, \&
  Band}]{Kaneko2006_166}
Kaneko, Y., Preece, R.~D., Briggs, M.~S., {et~al.} 2006, The Astrophysical
  Journal Supplement Series, 166, 298

\bibitem[{Klebesadel {et~al.}(1973)Klebesadel, Strong, \&
  Olson}]{Klebesadel1973_182}
Klebesadel, R.~W., Strong, I.~B., \& Olson, R.~A. 1973, The Astrophysical
  Journal, 182, L85

\bibitem[{{Kocevski}(2012)}]{Kocevski2012}
{Kocevski}, D. 2012, \apj, 747, 146

\bibitem[{Kouveliotou {et~al.}(1993)Kouveliotou, Meegan, Fishman, Bhat, Briggs,
  Koshut, Paciesas, \& Pendleton}]{Kouveliotou1993_413}
Kouveliotou, C., Meegan, C.~A., Fishman, G.~J., {et~al.} 1993, The
  Astrophysical Journal, 413, L101

\bibitem[{{Krisciunas} {et~al.}(2004){Krisciunas}, {Phillips}, \&
  {Suntzeff}}]{Krisciunas2004_602}
{Krisciunas}, K., {Phillips}, M.~M., \& {Suntzeff}, N.~B. 2004, \apjl, 602, L81

\bibitem[{{Kruehler} {et~al.}(2010){Kruehler}, {Greiner}, \&
  {Kann}}]{100728A_1.567_Kruehler2010GCN_14500}
{Kruehler}, T., {Greiner}, J., \& {Kann}, D.~A. 2010, GRB Coordinates Network,
  Circular Service, No.~14500, \#1 (2010), 14500

\bibitem[{{Kruehler} {et~al.}(2017){Kruehler}, {Schady}, {Greiner}, \&
  {Tanvir}}]{170214A_2.53_Kruehler2017GCN_20686}
{Kruehler}, T., {Schady}, P., {Greiner}, J., \& {Tanvir}, N.~R. 2017, GRB
  Coordinates Network, Circular Service, No.~20686, \#1 (2017), 20686

\bibitem[{{Kulkarni} {et~al.}(1998){Kulkarni}, {Frail}, {Wieringa}, {Ekers},
  {Sadler}, {Wark}, {Higdon}, {Phinney}, \& {Bloom}}]{Kulkarni1998}
{Kulkarni}, S.~R., {Frail}, D.~A., {Wieringa}, M.~H., {et~al.} 1998, \nat, 395,
  663

\bibitem[{{Levan} {et~al.}(2013){Levan}, {Cenko}, {Perley}, \&
  {Tanvir}}]{130427A_0.34_Levan2013GCN_14455}
{Levan}, A.~J., {Cenko}, S.~B., {Perley}, D.~A., \& {Tanvir}, N.~R. 2013, GRB
  Coordinates Network, Circular Service, No.~14455, \#1 (2013), 14455

\bibitem[{{Li}(2007)}]{Li2007}
{Li}, L.-X. 2007, \mnras, 379, L55

\bibitem[{{MacFadyen} \& {Woosley}(1999)}]{MacFadyen1999}
{MacFadyen}, A.~I., \& {Woosley}, S.~E. 1999, \apj, 524, 262

\bibitem[{{Malesani} {et~al.}(2009){Malesani}, {Goldoni}, {Fynbo}, {D'Elia},
  {Covino}, {Flores}, {Levan}, {Vergani}, \&
  {Wiersema}}]{090926A_2.1062_Malesani2009GCN_9942}
{Malesani}, D., {Goldoni}, P., {Fynbo}, J.~P.~U., {et~al.} 2009, GRB
  Coordinates Network, 9942

\bibitem[{{Meegan} {et~al.}(2009){Meegan}, {Lichti}, {Bhat}, {Bissaldi},
  {Briggs}, {Connaughton}, {Diehl}, {Fishman}, {Greiner}, {Hoover}, {van der
  Horst}, {von Kienlin}, {Kippen}, {Kouveliotou}, {McBreen}, {Paciesas},
  {Preece}, {Steinle}, {Wallace}, {Wilson}, \& {Wilson-Hodge}}]{Meegan2009_702}
{Meegan}, C., {Lichti}, G., {Bhat}, P.~N., {et~al.} 2009, The Astrophysical
  Journal, 702, 791

\bibitem[{{Paciesas} {et~al.}(2012){Paciesas}, {Meegan}, {von Kienlin}, {Bhat},
  {Bissaldi}, {Briggs}, {Burgess}, {Chaplin}, {Connaughton}, {Diehl},
  {Fishman}, {Fitzpatrick}, {Foley}, {Gibby}, {Giles}, {Goldstein}, {Greiner},
  {Gruber}, {Guiriec}, {van der Horst}, {Kippen}, {Kouveliotou}, {Lichti},
  {Lin}, {McBreen}, {Preece}, {Rau}, {Tierney}, \&
  {Wilson-Hodge}}]{2012ApJS..199...18P}
{Paciesas}, W.~S., {Meegan}, C.~A., {von Kienlin}, A., {et~al.} 2012, \apjs,
  199, 18

\bibitem[{{Pelassa} {et~al.}(2010){Pelassa}, {Preece}, {Piron}, {Omodei},
  {Guiriec}, {Fermi LAT Collaboration}, \& {Fermi GBM
  Collaboration}}]{Pelassa_2010}
{Pelassa}, V., {Preece}, R., {Piron}, F., {et~al.} 2010, ArXiv e-prints,
  arXiv:1002.2617

\bibitem[{{Perlmutter} \& {Schmidt}(2003)}]{Perlmutter_195}
{Perlmutter}, S., \& {Schmidt}, B.~P. 2003, in Lecture Notes in Physics, Berlin
  Springer Verlag, Vol. 598, Supernovae and Gamma-Ray Bursters, ed.
  K.~{Weiler}, 195--217

\bibitem[{{Perlmutter} {et~al.}(1999){Perlmutter}, {Aldering}, {Goldhaber},
  {Knop}, {Nugent}, {Castro}, {Deustua}, {Fabbro}, {Goobar}, {Groom}, {Hook},
  {Kim}, {Kim}, {Lee}, {Nunes}, {Pain}, {Pennypacker}, {Quimby}, {Lidman},
  {Ellis}, {Irwin}, {McMahon}, {Ruiz-Lapuente}, {Walton}, {Schaefer}, {Boyle},
  {Filippenko}, {Matheson}, {Fruchter}, {Panagia}, {Newberg}, {Couch}, \&
  {Project}}]{Perlmutter1999_517}
{Perlmutter}, S., {Aldering}, G., {Goldhaber}, G., {et~al.} 1999, The
  Astrophysical Journal, 517, 565

\bibitem[{{Petrosian} {et~al.}(2015){Petrosian}, {Kitanidis}, \&
  {Kocevski}}]{2015ApJ...806...44P}
{Petrosian}, V., {Kitanidis}, E., \& {Kocevski}, D. 2015, \apj, 806, 44

\bibitem[{{Planck Collaboration} {et~al.}(2018){Planck Collaboration},
  {Aghanim}, {Akrami}, {Ashdown}, {Aumont}, {Baccigalupi}, {Ballardini},
  {Banday}, {Barreiro}, {Bartolo}, {Basak}, {Battye}, {Benabed}, {Bernard},
  {Bersanelli}, {Bielewicz}, {Bock}, {Bond}, {Borrill}, {Bouchet}, {Boulanger},
  {Bucher}, {Burigana}, {Butler}, {Calabrese}, {Cardoso}, {Carron},
  {Challinor}, {Chiang}, {Chluba}, {Colombo}, {Combet}, {Contreras}, {Crill},
  {Cuttaia}, {de Bernardis}, {de Zotti}, {Delabrouille}, {Delouis}, {Di
  Valentino}, {Diego}, {Dor{\'e}}, {Douspis}, {Ducout}, {Dupac}, {Dusini},
  {Efstathiou}, {Elsner}, {En{\ss}lin}, {Eriksen}, {Fantaye}, {Farhang},
  {Fergusson}, {Fernandez-Cobos}, {Finelli}, {Forastieri}, {Frailis},
  {Franceschi}, {Frolov}, {Galeotta}, {Galli}, {Ganga}, {G{\'e}nova-Santos},
  {Gerbino}, {Ghosh}, {Gonz{\'a}lez-Nuevo}, {G{\'o}rski}, {Gratton},
  {Gruppuso}, {Gudmundsson}, {Hamann}, {Handley}, {Herranz}, {Hivon}, {Huang},
  {Jaffe}, {Jones}, {Karakci}, {Keih{\"a}nen}, {Keskitalo}, {Kiiveri}, {Kim},
  {Kisner}, {Knox}, {Krachmalnicoff}, {Kunz}, {Kurki-Suonio}, {Lagache},
  {Lamarre}, {Lasenby}, {Lattanzi}, {Lawrence}, {Le Jeune}, {Lemos},
  {Lesgourgues}, {Levrier}, {Lewis}, {Liguori}, {Lilje}, {Lilley}, {Lindholm},
  {L{\'o}pez-Caniego}, {Lubin}, {Ma}, {Mac{\'{\i}}as-P{\'e}rez}, {Maggio},
  {Maino}, {Mandolesi}, {Mangilli}, {Marcos-Caballero}, {Maris}, {Martin},
  {Martinelli}, {Mart{\'{\i}}nez-Gonz{\'a}lez}, {Matarrese}, {Mauri}, {McEwen},
  {Meinhold}, {Melchiorri}, {Mennella}, {Migliaccio}, {Millea}, {Mitra},
  {Miville-Desch{\^e}nes}, {Molinari}, {Montier}, {Morgante}, {Moss}, {Natoli},
  {N{\o}rgaard-Nielsen}, {Pagano}, {Paoletti}, {Partridge}, {Patanchon},
  {Peiris}, {Perrotta}, {Pettorino}, {Piacentini}, {Polastri}, {Polenta},
  {Puget}, {Rachen}, {Reinecke}, {Remazeilles}, {Renzi}, {Rocha}, {Rosset},
  {Roudier}, {Rubi{\~n}o-Mart{\'{\i}}n}, {Ruiz-Granados}, {Salvati}, {Sandri},
  {Savelainen}, {Scott}, {Shellard}, {Sirignano}, {Sirri}, {Spencer},
  {Sunyaev}, {Suur-Uski}, {Tauber}, {Tavagnacco}, {Tenti}, {Toffolatti},
  {Tomasi}, {Trombetti}, {Valenziano}, {Valiviita}, {Van Tent}, {Vibert},
  {Vielva}, {Villa}, {Vittorio}, {Wandelt}, {Wehus}, {White}, {White},
  {Zacchei}, \& {Zonca}}]{Planck2018}
{Planck Collaboration}, {Aghanim}, N., {Akrami}, Y., {et~al.} 2018, arXiv
  e-prints, arXiv:1807.06209

\bibitem[{{Pugliese} {et~al.}(2015){Pugliese}, {Xu}, {Tanvir}, {Wiersema},
  {Fynbo}, {Milvang-Jensen}, \& {D'Elia}}]{150403A_2.06_Pugliese2015GCN_17672}
{Pugliese}, V., {Xu}, D., {Tanvir}, N.~R., {et~al.} 2015, GRB Coordinates
  Network, Circular Service, No.~17672, \#1 (2015), 17672

\bibitem[{{Rau} {et~al.}(2009){Rau}, {McBreen}, \&
  {Kruehler}}]{GRB090510_0.903_Rau2009GCN_9353}
{Rau}, A., {McBreen}, S., \& {Kruehler}, T. 2009, GRB Coordinates Network, 9353

\bibitem[{{Ravasio} {et~al.}(2018){Ravasio}, {Oganesyan}, {Ghirlanda}, {Nava},
  {Ghisellini}, {Pescalli}, \& {Celotti}}]{Ravasio2018_613}
{Ravasio}, M.~E., {Oganesyan}, G., {Ghirlanda}, G., {et~al.} 2018, \aap, 613,
  A16

\bibitem[{Riess {et~al.}(1998)Riess, Filippenko, Challis, Clocchiatti, Diercks,
  Garnavich, Gilliland, Hogan, Jha, Kirshner, {et~al.}}]{Riess1998_116}
Riess, A.~G., Filippenko, A.~V., Challis, P., {et~al.} 1998, The Astronomical
  Journal, 116, 1009

\bibitem[{{Ryde}(1999)}]{Ryde1999_39}
{Ryde}, F. 1999, Astrophysical Letters and Communications, 39, 281

\bibitem[{{Sanchez-Ramirez} {et~al.}(2013){Sanchez-Ramirez}, {Gorosabel},
  {Castro-Tirado}, {Cepa}, \&
  {Gomez-Velarde}}]{130518A_2.49_Sanchez-Ramirez2013GCN_14685}
{Sanchez-Ramirez}, R., {Gorosabel}, J., {Castro-Tirado}, A.~J., {Cepa}, J., \&
  {Gomez-Velarde}, G. 2013, GRB Coordinates Network, Circular Service,
  No.~14685, \#1 (2013), 14685

\bibitem[{{Soderberg} {et~al.}(2006){Soderberg}, {Kulkarni}, {Nakar}, {Berger},
  {Cameron}, {Fox}, {Frail}, {Gal-Yam}, {Sari}, {Cenko}, {Kasliwal},
  {Chevalier}, {Piran}, {Price}, {Schmidt}, {Pooley}, {Moon}, {Penprase},
  {Ofek}, {Rau}, {Gehrels}, {Nousek}, {Burrows}, {Persson}, \&
  {McCarthy}}]{Soderberg2006}
{Soderberg}, A.~M., {Kulkarni}, S.~R., {Nakar}, E., {et~al.} 2006, \nat, 442,
  1014

\bibitem[{{Stanek} {et~al.}(2003){Stanek}, {Matheson}, {Garnavich}, {Martini},
  {Berlind}, {Caldwell}, {Challis}, {Brown}, {Schild}, {Krisciunas}, {Calkins},
  {Lee}, {Hathi}, {Jansen}, {Windhorst}, {Echevarria}, {Eisenstein}, {Pindor},
  {Olszewski}, {Harding}, {Holland}, \& {Bersier}}]{Stanek2003}
{Stanek}, K.~Z., {Matheson}, T., {Garnavich}, P.~M., {et~al.} 2003, \apjl, 591,
  L17

\bibitem[{{Suzuki} {et~al.}(2012){Suzuki}, {Rubin}, {Lidman}, {Aldering},
  {Amanullah}, {Barbary}, {Barrientos}, {Botyanszki}, {Brodwin}, {Connolly},
  {Dawson}, {Dey}, {Doi}, {Donahue}, {Deustua}, {Eisenhardt}, {Ellingson},
  {Faccioli}, {Fadeyev}, {Fakhouri}, {Fruchter}, {Gilbank}, {Gladders},
  {Goldhaber}, {Gonzalez}, {Goobar}, {Gude}, {Hattori}, {Hoekstra}, {Hsiao},
  {Huang}, {Ihara}, {Jee}, {Johnston}, {Kashikawa}, {Koester}, {Konishi},
  {Kowalski}, {Linder}, {Lubin}, {Melbourne}, {Meyers}, {Morokuma}, {Munshi},
  {Mullis}, {Oda}, {Panagia}, {Perlmutter}, {Postman}, {Pritchard}, {Rhodes},
  {Ripoche}, {Rosati}, {Schlegel}, {Spadafora}, {Stanford}, {Stanishev},
  {Stern}, {Strovink}, {Takanashi}, {Tokita}, {Wagner}, {Wang}, {Yasuda},
  {Yee}, \& {Supernova Cosmology Project}}]{Suzuki2012_746}
{Suzuki}, N., {Rubin}, D., {Lidman}, C., {et~al.} 2012, \apj, 746, 85

\bibitem[{{Tanvir} {et~al.}(2016){Tanvir}, {Levan}, {Cenko}, {Perley},
  {Cucchiara}, {Roth}, {Wiersema}, {Fruchter}, \&
  {Laskar}}]{160509A_1.17_Tanvir2016GCN_19419}
{Tanvir}, N.~R., {Levan}, A.~J., {Cenko}, S.~B., {et~al.} 2016, GRB Coordinates
  Network, Circular Service, No.~19419, \#1 (2016), 19419

\bibitem[{{Tsvetkova} {et~al.}(2017){Tsvetkova}, {Frederiks}, {Golenetskii},
  {Lysenko}, {Oleynik}, {Pal'shin}, {Svinkin}, {Ulanov}, {Cline}, {Hurley}, \&
  {Aptekar}}]{2017ApJ...850..161T}
{Tsvetkova}, A., {Frederiks}, D., {Golenetskii}, S., {et~al.} 2017, \apj, 850,
  161

\bibitem[{{Wang} {et~al.}(2016){Wang}, {Wang}, {Cheng}, \&
  {Dai}}]{Wang2016_585}
{Wang}, J.~S., {Wang}, F.~Y., {Cheng}, K.~S., \& {Dai}, Z.~G. 2016, \aap, 585,
  A68

\bibitem[{{Wiersema} {et~al.}(2009){Wiersema}, {Tanvir}, {Cucchiara}, {Levan},
  \& {Fox}}]{091208B_1.063_Wiersema2009GCN_10263}
{Wiersema}, K., {Tanvir}, N.~R., {Cucchiara}, A., {Levan}, A.~J., \& {Fox}, D.
  2009, GRB Coordinates Network, Circular Service, No.~10263, \#1 (2009), 10263

\bibitem[{Wilks(1938)}]{Wilks1938_9}
Wilks, S.~S. 1938, The Annals of Mathematical Statistics, 9, 60

\bibitem[{{Wood-Vasey} {et~al.}(2008){Wood-Vasey}, {Friedman}, {Bloom},
  {Hicken}, {Modjaz}, {Kirshner}, {Starr}, {Blake}, {Falco}, {Szentgyorgyi},
  {Challis}, {Blondin}, {Mandel}, \& {Rest}}]{Wood_Vasey2008_689}
{Wood-Vasey}, W.~M., {Friedman}, A.~S., {Bloom}, J.~S., {et~al.} 2008, \apj,
  689, 377

\bibitem[{{Xu} {et~al.}(2014){Xu}, {Levan}, {Fynbo}, {Tanvir}, {D'Elia}, \&
  {Malesani}}]{141028A_2.33_Xu2014GCN_16983}
{Xu}, D., {Levan}, A.~J., {Fynbo}, J.~P.~U., {et~al.} 2014, GRB Coordinates
  Network, Circular Service, No.~16983, \#1 (2014), 16983

\bibitem[{{Xu} {et~al.}(2016){Xu}, {Malesani}, {Fynbo}, {Tanvir}, {Levan}, \&
  {Perley}}]{160625B_1.406_Xu2016GCN_19600}
{Xu}, D., {Malesani}, D., {Fynbo}, J.~P.~U., {et~al.} 2016, GRB Coordinates
  Network, Circular Service, No.~19600, \#1 (2016), 19600

\bibitem[{{Yonetoku} {et~al.}(2004){Yonetoku}, {Murakami}, {Nakamura},
  {Yamazaki}, {Inoue}, \& {Ioka}}]{Yonetoku2004_609}
{Yonetoku}, D., {Murakami}, T., {Nakamura}, T., {et~al.} 2004, \apj, 609, 935

\bibitem[{{Yu} {et~al.}(2016){Yu}, {Preece}, {Greiner}, {Narayana Bhat},
  {Bissaldi}, {Briggs}, {Cleveland}, {Connaughton}, {Goldstein}, {von Kienlin},
  {Kouveliotou}, {Mailyan}, {Meegan}, {Paciesas}, {Rau}, {Roberts}, {Veres},
  {Wilson-Hodge}, {Zhang}, \& {van Eerten}}]{2016A&A...588A.135Y}
{Yu}, H.-F., {Preece}, R.~D., {Greiner}, J., {et~al.} 2016, \aap, 588, A135

\end{thebibliography}
\bibliographystyle{apj}
\end{document}